\documentclass[pra,amsmath,amssymb,twocolumn]{revtex4-2}

\usepackage{graphicx}
\usepackage{dcolumn}
\usepackage{bm}

\usepackage[T1]{fontenc}

\usepackage{dsfont}
\usepackage{appendix}
\usepackage{subcaption}
\usepackage{graphicx}
\usepackage{chemarrow}
\usepackage{amsthm,amssymb,amsmath}
\usepackage{array}
\usepackage{hyperref}
\usepackage{xcolor}

\begin{document}

\title{Frame-filtered ghost imaging with a SPAD array used both as a multiple "bucket" detector and an imaging camera} 

\author{V. S. Starovoitov}
\affiliation{B.I.Stepanov Institute of Physics, NAS of Belarus, Nezavisimosti ave. 68, 220072 Minsk, Belarus}

\author{V. N. Chizhevsky}
\affiliation{B.I.Stepanov Institute of Physics, NAS of Belarus, Nezavisimosti ave. 68, 220072 Minsk, Belarus}

\author{D. Mogilevtsev}
\affiliation{B.I.Stepanov Institute of Physics, NAS of Belarus, Nezavisimosti ave. 68, 220072 Minsk, Belarus}

\author{A. Smaliakou}
\affiliation{B.I.Stepanov Institute of Physics, NAS of Belarus, Nezavisimosti ave. 68, 220072 Minsk, Belarus}

\author{M. Perenzoni}
\affiliation{Fondazione Bruno Kessler FBK, 38122 Trento, Italy}
\affiliation{now at Sony Semiconductor Solutions Europe, Trento, Italy}

\author{ L. Gasparini}
\affiliation{Fondazione Bruno Kessler FBK, 38122 Trento, Italy}

\author{D. B. Horoshko}
\affiliation{Univ. Lille, PhLAM - Physique des Lasers Atomes et Molecules, F-59000 Lille, France}

\author{S. Kilin}
\affiliation{B.I.Stepanov Institute of Physics, NAS of Belarus, Nezavisimosti ave. 68, 220072 Minsk, Belarus}

\date{April 2024}

\begin{abstract}
An approach to ghost imaging with a single SPAD array used simultaneously as a several-pixel "bucket" detector and an imaging camera is described. The key points of the approach are filtering data frames used for ghost-image reconstruction by the number of per-frame counts and superposing correlation images obtained for different "bucket" pixels. The imaging is performed in an experiment with a pseudo-thermal light source where the light intensity is so low that the dark counts have a noticeable effect on imaging. We demonstrate that the approach is capable to significantly reduce the destructive effect of dark counts on the ghost image and improve image contrast, spatial resolution, and image similarity to a  reference image.  
\end{abstract}

\maketitle

\section{Introduction}

Ghost imaging (GI) is already a mature research field dating back nearly a quarter of century. Since the first demonstration for GI with twin-photons \cite{PhysRevA.52.R3429}  and subsequent GI realization with classically correlated light \cite{PhysRevLett.89.113601,PhysRevLett.94.063601,doi:10.1080/09500340500147240}, a large corpus of experimental and theoretical works has appeared addressing particular aspects of GI and its applications for imaging/sensing (see, for example, review works \cite{Erkmen:10,Shapiro2012ThePO,doi:10.1098/rsta.2016.0233,Simon2017,Gibson:20,10.1117/12.2662291}).  The major practical appeal of GI {is a} possibility to use for image detection a maximally simple bucket detector unable to produce an image by itself. One has to use high-resolution detection only with the reference beam (a part of the correlated light not interacting with the object). Moreover, even correlations of the bucket-detector photo-current with the reference current  are able to provide data sufficient for building a correlation image \cite{PhysRevA.78.061802} (this possibility gave rise to a rapidly developing "single-pixel" imaging technique with such possible applications as detection of gas leaks and sensors for situational awareness for vehicles \cite{2019NaPho..13...13E}). 

One of the most perspective GI ways to enhance imaging sensitivity and account for non-classical light correlations is to apply recently developed single photon avalanche diode (SPAD) arrays for the reference-beam detection \cite{doi:10.1098/rsta.2016.0233}.   Currently, the SPAD arrays operated in various detection regimes (such as time-stamping, counting, and gating) have reached a high technological maturity \cite{Bruschini}, featuring observation rates up to 1 MHz, 200 ps time resolution \cite{9142240}, close to 100 $\%$ fill factor with megapixel spatial resolution \cite{9720605}. 
SPAD cameras have already found a widespread use in imaging, for example, in near-infrared fluorescence lifetime imaging \cite{Smith:22}, super-resolving quantum imaging \cite{defienne,Unternahrer:18}, demonstrating of EPR inequality violation by measurement of photon correlations \cite{Eckmann:20}, correlation plenoptic imaging \cite{app11146414}, Hong-Ou-Mandel interference microscopy \cite{ndagano}, LIDAR applications \cite{Zhao:22}. 
In these applications, the image is reconstructed from data acquired in fixed temporal windows, hereafter referred to as frames.

The dark counts (DC) are an inevitable effect that limits the sensitivity of SPAD-based imaging. The DC are caused by intrinsic generation of carriers within the detector in absence of illumination. The DC are not correlated with the photocounts, but they, however, cannot be separated from them.  The higher the rate of DC in a pixel, the lower the correlation with the light for the events it detects. 
In large-size detector arrays, a significant percentage of detector-pixels (such pixels are called “hot” pixels) could show a DC rate much higher than the median DC rate of the array, sometimes even exceeding the typical detected signal. In correlation-based imaging applications, the destructive action of DC cannot be reduced by data averaging over time. 
Eventually, the effect of dark counts on measurement may be minimized by disabling noisy pixels \cite{6642135,8700491,Xu:17}.

In this work we use the SPAD array for a ghost imaging modality that makes a step further toward full-size quantum correlation imaging with aim to improve image quality. Namely, instead of a single-pixel bucket detector we use several pixels of the imaging SPAD camera (as in \cite{doi:10.1080/09500340500147240,Magnitskiy:22}) and perform averaging over a ghost images obtained for each "bucket" pixel. Also, we propose procedures allowing to deal with intrinsic shortcomings of SPADs in the low-light regime. The procedures are frame-filtering on the number of counts per frame and algorithms for dealing with dark counts in-homogeneously distributed over the SPAD pixels.  We demonstrate that for a standard pseudo-thermal source of correlated photons (namely, a rotating ground-glass disk) such an approach can significantly rise the ghost-image quality. 

It should be emphasized that key points of our approach, namely, filtering frames with respect to a specific number of counts and averaging over several simultaneously obtained intensity correlation images look to be rather general and applicable not only for different ghost imaging modalities, but also for generic correlation imaging with light on the few-photon level, first of all, for quantum microscopy/sensing applications.

The outline of the paper is as follows. In the second Section we describe the ghost imaging set-up, the object and the SPAD camera used for imaging. In the third Section we outline the data post-processing procedure, give the definition of the ghost image for the case of multiple bucket detectors and define two measures of image quality: the contrast-to-noise ratio and the correlation coefficient to the reference image. In the fourth Section the ghost image inference and quantification of the image contrast and quality are discussed. Here we demonstrate ghost imaging with filtered and unfiltered data, and also compare results averaged over multiple "bucket" pixels and just a single "bucket" pixel. In the Appendix, the influence of dark counts on the ghost image  is discussed.

\begin{figure}
    \centering
    \includegraphics[width=0.99\linewidth]{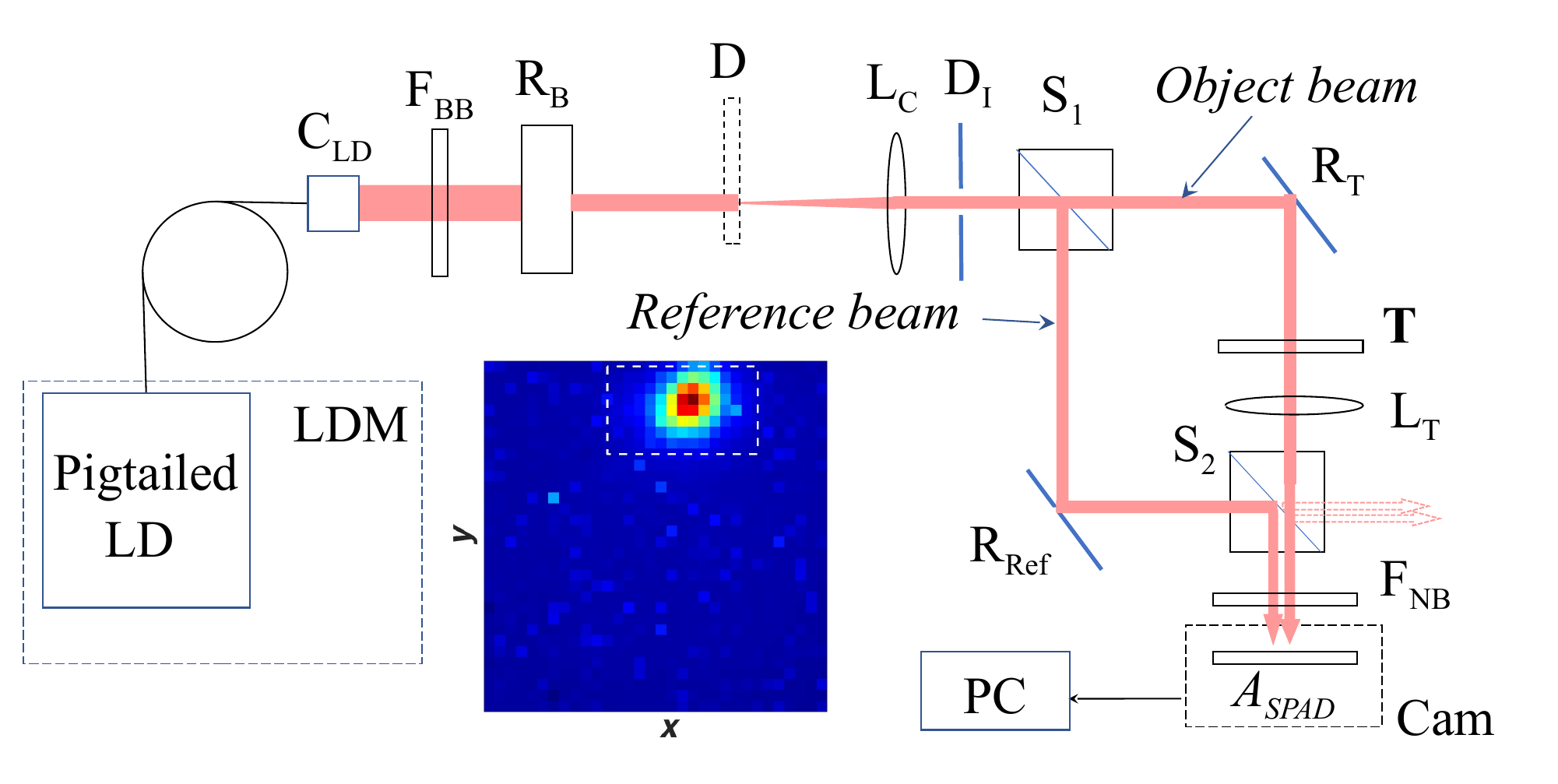}
    \caption{{The scheme of the ghost imaging experiment: ($LD$) laser diode; ($LDM$) laser diode mount; $C_{LD}$ fiber collimator; ($F_{BB}$) broadband filter; ($R_B$) beam reducer; ($D$) rotating ground-glass disk;     ($L_C$, $L_T$) lenses; ($D_I$) iris diaphragm; ($S_1$, $S_2$) beam splitters; ($R_{Ref}$, $R_{T}$) reflectors; ($T$) target; ($F_{NB}$) narrow-band filter; ($A_{SPAD}$) SPAD array; ($Cam$) camera "SuperEllen"; ($PC$) computer. The inset shows how the SPAD array is illuminated by the object and reference beams. The dashed line in the inset shows boundaries of the “Flare” area; the "bucket" pixels are there. }}
    \label{fig1}
\end{figure}

\section{Setup}

\subsection{Optical layout}
In the work, the GI is performed by an experimental way using correlated light beams produced with help of a laser and a rotating ground-glass disk. The initial light beam is generated by a CW-operated laser source. The experimental setup is shown in Fig~\ref{fig1}. The source is a fiber-pigtailed single-transverse-mode Fabry-Perot laser diode (LDS-670-FP-5-U-3-SMP04-FU-CW-0.5, LaserCom LLC) operated near a wavelength $\lambda\approx671$ nm with an output power less than 8 mW. The diode is thermally stabilized inside a laser mount (Thorlabs LDM9T) and operates at constant current in near-threshold mode (power supply Thorlabs LDC201CU).  The laser beam is output from a PM-fiber and a fiber collimator ($C_{LD}$, efficient focus length of 7.5 mm) mounted in a Thorlabs PAF2-7B fiber-port. This beam is collimated. Its cross section has a Gaussian profile with a diameter of approximately 1.4 mm. The beam is directed through a broadband filter $F_{BB}$ and a free-space beam reducer ($R_B$, of about 3 times reduction factor) towards a rotating ground-glass disk (D). The distances $|C_{LD}D|$ and $|R_{B}D|$ are about 100 mm and 20 mm, respectively. The diameter of spot illuminated by this beam on the rotating disk is $d_d \approx 0.47$ mm. Disc D has a matte surface with a random grained transparent pattern that scatters the input optical beam over a wide solid angle. The luminous laser spot on the surface of disk D is a pseudo-thermal source of correlated photons that form random structure of light spots (also called speckles) in space. The distance between the optical axis and the axis of disk rotation is 52 mm. The disk rotates at a frequency of 10 Hz. A lens ($L_C$, focal length $F_{LC} = 80$ mm) positioned at a distance $|DL_C| \approx F_{LC}$ collects photons scattered by the disk, collimates and directs them through an iris diaphragm ($D_I$, $|DD_I| \approx 20$ mm) towards a beam splitter ($S_1$, $|DS_1| \approx 55$ mm). 

Splitter $S_1$ divides the collimated speckle beam into two arms: object and reference. In the reference arm, the beam is directed through a reflector ($R_{Ref}$, $|S_1R_{Ref}| \approx 105$ mm) and another beam splitter ($S_{2}$, $|R_TS_{2}| \approx 85$ mm) towards a SPAD array ($A_{SPAD}$) mounted in a camera (Cam). In the object arm, the beam is directed by a reflector ($R_T$, $|S_1R_T| \approx 85$ mm) to a target (T). A lens ($L_T$, focus length $F_{LT} = 67$ mm) collects the light transmitted by target T and focuses it (through beam splitter $S_2$) on array $A_{SPAD}$ ($|L_TA_{SPAD}| \approx F_{LT}$). Target T is positioned as close as possible to splitter $S_2$. The distances $|TL_T|$ and $|L_TS_2|$ are 13 and 15 mm, respectively. A 20 nm narrowband filter ($F_{NB}$) is located in front of the camera for suppressing influence of stray light. The distance $|S_2A_{SPAD}| \approx 50$ mm. The beam splitters $S_1$ and $S_2$ are 50:50 non-polarizing beamsplitter cubes BS011 (Thorlabs). The reflectors $R_{Ref}$ and $R_T$ are Thorlabs right-angle prism FS910L-B. 

Splitter $S_2$ directs photons from both the reference and object arms to array $A_{SPAD}$. The reference beam uniformly fills the entire surface of array $A_{SPAD}$. The object arm is adjusted so that the light spot focused by lens $L_T$ on the surface of array $A_{SPAD}$ falls on the upper right part of $A_{SPAD}$ (see the inset in Fig.~\ref{fig1}(a)). The diameter of this spot is $d_T = d_d F_{LT}/F_C \approx 0.39$ mm. The central part of this spot (referred to as “bucket”) is intended to be used as a bucket detector. A larger area (an area limited by dashed lines in the inset in Fig.~\ref{fig1}(a)) around the object-beam spot is considered as a flare area, unsuitable for image reconstruction. The remaining part of array $A_{SPAD}$ (called “Ghost”) lighted only by the reference beam is reserved for ghost image reconstruction. A similar arrangement of having a "bucket" and "ghost" images in the region of just a single camera (albeit a CCD one) was used in an earlier work on ghost imaging with quasi-classical field \cite{doi:10.1080/09500340500147240}. Recently, a similar set-up was used in works on ghost polarimetry \cite{Magnitskiy:20,Magnitskiy:21,Magnitskiy:22}. 

\subsection{Imaged object and SPAD array}
\begin{figure}
    \centering
    \includegraphics[width=0.99\linewidth]{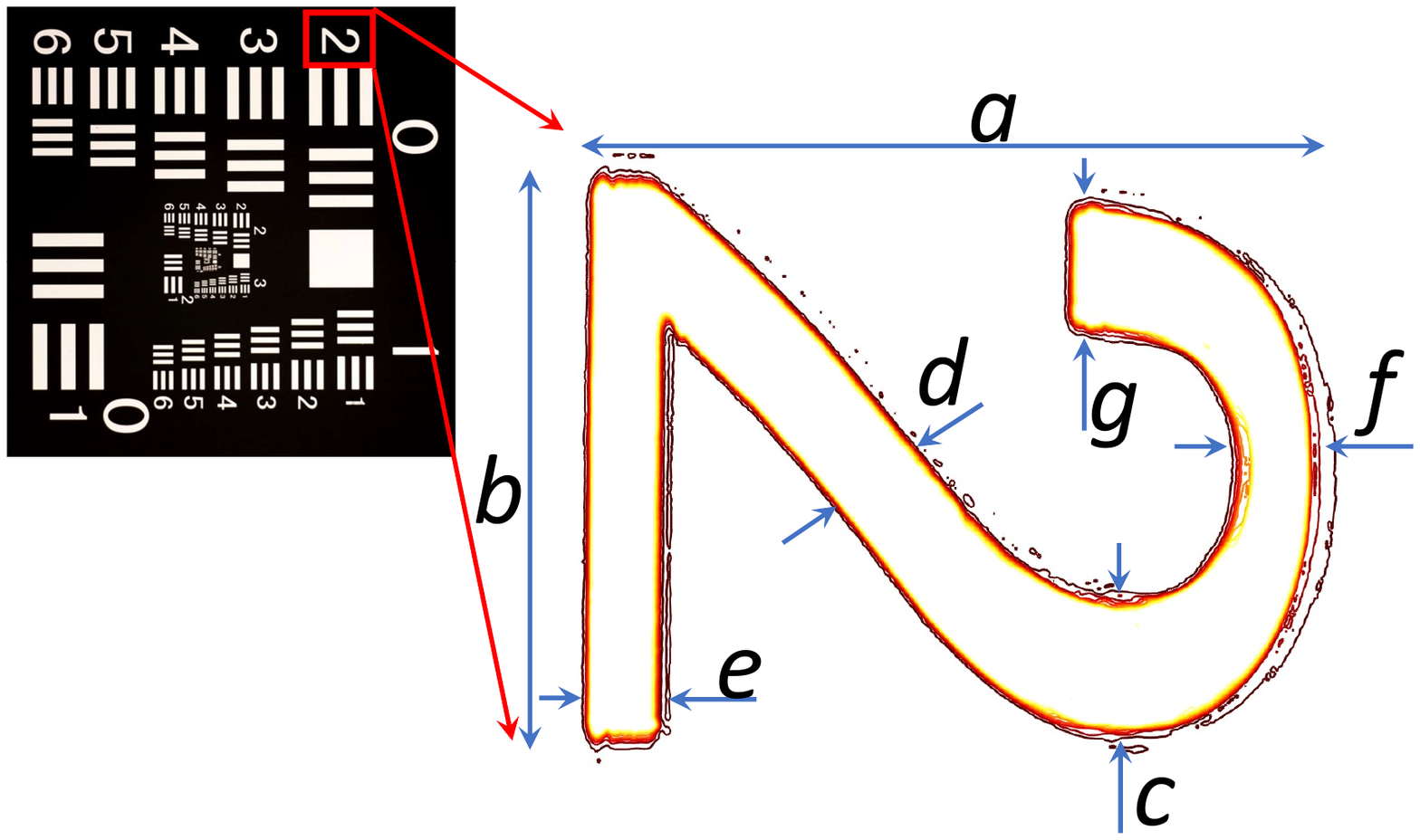}
    \caption{Resolution test target (left) and high-resolution contour plot of the object with sizes (right): \protect{$a\approx 1.3$ mm; $b\approx 0.94$ mm; $c\approx 0.20$ mm; $d\approx 0.16$ mm; $e\approx 0.116$ mm; $f\approx 0.12$ mm; $g\approx 0.19$ mm.}}
    \label{fig:Target}
\end{figure}

The imaged object is a part of a 1951 USAF resolution test target (negative chrome-on-glass version, see Fig.~\ref{fig:Target}) discussed in detail in \cite{Target}. We reproduce the images for the digit template “2” (Element 2 of Group 0) of the test target. In order to illuminate the template ”2” selectively from other target patterns, the light beams are restricted by diaphragm $D_I$. The height and width of the selected template are 1.3 and 0.94 mm. The font-line thickness for this object is a value varying from 0.11 to 0.20 mm. The image of the selected digit template is shown in Fig.~\ref{fig:Target}. 

The SPAD camera (Cam) is based on the "SuperEllen" sensor, specifically developed by Fondazione Bruno Kessler to target quantum imaging applications  \cite{Unternahrer:18,9142240} and implemented recently also for ghost imaging \cite{Pitsch:23,Gili:23}.
The array consists of $32\times32=1024$ pixels with a pitch of $\delta_s = 44.54 \mu m$ and a total sensitive area of $1.4\times 1.4 mm^2$ manufactured in 150 nm CMOS standard technology. The pixels are addressed by their linear index $k\in\{1,\ldots,1024\}$ such that $k=y_k+32 (x_k-1)$ where $x_k,y_k\in\{1,\ldots,32\}$ are the discrete Cartesian coordinates of pixels. For every pixel there is a  time-to-digital converter (TDC) which timestamps the first count with up to 204 ps resolution within a frame of exposure time  of up to 100 ns. Using on-chip features for reading empty rows or frames, a frequency of exposure repetition (i.e. an observation rate) of 800 kHz is achieved leading to a measurement duty-cycle of 3.6$\%$. The photon detection efficiency reaches $5\%$ at 400 nm and $0.8\%$ at $810$ nm. We did not specifically measure the efficiency at the used wavelength $671$ nm. However, we estimate it as being between $1.5\%$ and $2 \%$. 
The median dark count rate per pixel is approximately $0.7$ kHz over the whole pixel population.  The timestamped counts are registered in a frame-by-frame way. 

In our experiment, each frame is a result of light exposition during the time of $45$ ns. As it is natural to expect with our pseudo-thermal light source,  there are time-correlation between frames. We have addressed these correlations in our recent work \cite{2023JApSp..90..377S}. However, their presence does not affect current consideration. 
The data related to an individual frame is converted into a compressed sparse structure
\begin{equation}
\mathcal{D}_f:=\{f,n_f,\{x_k^{(i)},y_k^{(i)},\tau^{(i)}\}_{i=1,\ldots,n_f}\},
\label{FrameSet}
\end{equation}
where $f$ is the serial number of frame repetition; $n_f$ is the total number of events (counts) 
detected by all pixels in frame $f$; $x_k^{(i)}$, $y_k^{(i)}$ and $\tau^{(i)}$ are the discrete Cartesian coordinates and the detection time for the $i$-th event registered in frame $f$. The registered data is collected and transferred to a computer (PC) as a stream of multiple frames 
$\mathcal{F}=\{\mathcal{D}_f\big|n_f\geq1\}$. 
The frames $\{\mathcal{D}_f|n_f=0\}$ (frames for which no events were detected) are not included in the stream for accelerating the data transfer process.

\section{Post-processing}
\subsection{Datasets and correlation functions}
For ghost imaging both unfiltered and filtered datasets will be used. The unfiltered  dataset  $F_{0}=\bigl\{\mathcal{D}_f\big|n_f\geq0\bigr\}$,
contains all frames registered in the experiment (including the frames of original stream $\mathcal{F}$  and the frames $\{\mathcal{D}_f|n_f=0\}$ omitted at data transfer. The total number of frames in this dataset is  $N_{0}$. 

The  frame-filtered datasets  
\begin{equation}
F(n^*)=F_{0}\big|_{n_f=n^*}=\bigl\{ \mathcal{D}_f\big|n_f=n^*\bigr\}
    \label{FilterData}
\end{equation}
are non-overlapping sub-sets of $F_{0}$ with fixed number of counts $n^*$ in frame. The ratio 
\begin{equation}
p(n_f=n^*)={N(n^*)}/{N_{0}}
    \label{ProbDistr}
\end{equation}
   is normalised distribution of filtered sub-sets number $N(n^*)$.

\begin{figure}
    \centering
    \includegraphics[width=0.99\linewidth]{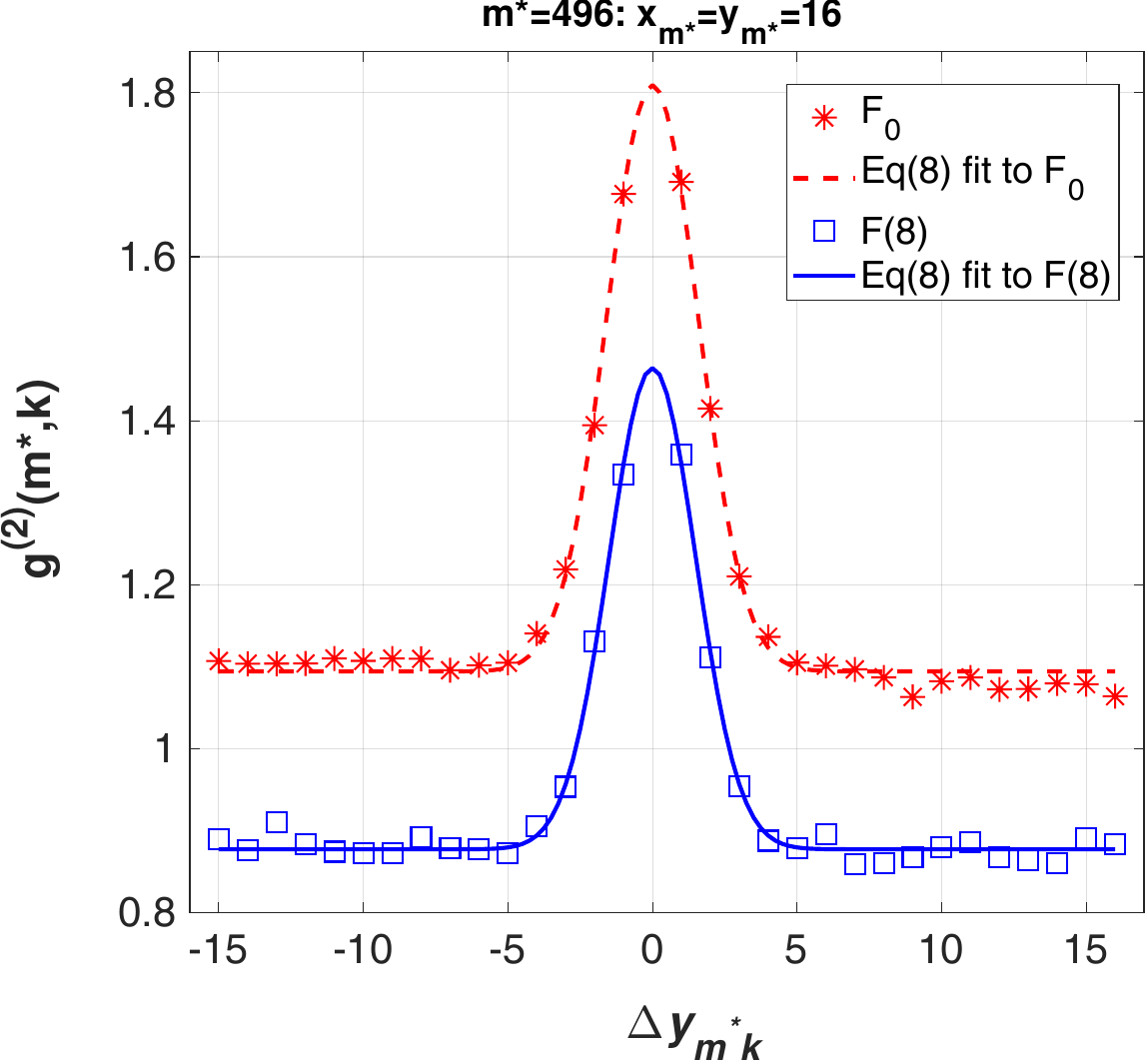}
    \caption{ Functions \protect{$g^{(2)}(m^{*},k)$} depending on the distance $\Delta y_{m^{*}k}$ for the pixel \protect{$m^*=496$ ($x_{m^*}=y_{m^*}=16$}). The functions are acquired from a frame-filtered dataset $F(8)$ (blue squares) and unfiltered data $F_0$ (red stars). The blue solid ($F(8)$) and red dashed ($F_0$) lines indicate approximation  (\ref{Shape}) to the acquired functions. }
    \label{fig:CorrFun}
\end{figure}

The prepared dataset is used to calculate  both a per-frame distribution of counts and their pairwise correlations\cite{2023JApSp..90..377S}:
\begin{eqnarray}
G^{(1)}(k)=\langle P_k\rangle_{\{f\}}, 
\quad
G^{(2)}(m,k)=\langle P_mP_k\rangle_{\{f\}},
\label{G12}
\end{eqnarray}
where $k$ and $m$ are linear pixel indices, and $\langle \ldots \rangle_{\{f\}}$ denotes averaging over the dataset frames.  The quantities $P_k$ in \eqref{G12} are random binaries taking the value 1 for a count registration at the pixel $k$ or zero in its absence. In the following consideration we use the normalized second-order correlation function 
\begin{equation}
g^{(2)}(m,k)={G^{(2)}(m,k)}/{G^{(1)}(m)G^{(1)}(k)},
\label{g2}
\end{equation}
conditioned by the choice of dataset. 

The ghost imaging implies knowledge of parameters (spatial resolution, correlation strength and background) that specify the correlation  over SPAD pixels for the applied light beams. The correlation $g^{(2)}(m^{*},k)$ between a “probing” low-noise pixel $m=m^{*}$, selected inside the "ghost" area (Fig.~\ref{fig1}),  and all other pixels of the SPAD array
is well approximated by the Gaussian
\begin{equation}
g^{(2)}(m^{*},k)\approx A_{m^{*}}\exp\left\{-\frac{|r_{m^{*}k}|^2}{2B_{m^{*}}^2}\right\}+C_{m^{*}},
\label{Shape}
\end{equation}
with the distance $|r_{m^{*}k}|=\sqrt{\Delta x_{m^{*}k}^2+\Delta y_{m^{*}k}^2 }$ between $k$-th and $m^{*}$-th pixels. 
Typical functions $g^{(2)}(m^{*},k)$ depending on the distance $\Delta y_{m^{*}k}$ acquired from different datasets for a '"probing" pixel $m^*$ (the pixel is located in the middle of the SPAD array) are demonstrated in Fig.  (\ref{fig:CorrFun}).

\subsection{Ghost image and its quality}
In the work, we define the ghost image as a weighted sum of several correlation functions
\begin{equation}
 \mathcal{I}'_k={\sum\limits_{m\in \{B\}} G^{(1)}(m)g^{(2)}(m,k)}/{\sum\limits_{m\in \{B\}} G^{(1)}(m)}.
    \label{ghost0}
\end{equation}
 The pixels labeled by $m$ are "bucket" pixels belonging to the set $\{B\}$. These pixels are chosen as located near the center of the object-beam spot on the SPAD area. The image pixels labeled by $k$ are in the "ghost" area. 

The scaling factor and background of the image $\mathcal{I}'_k$ are dataset dependent. In order to minimize such an uncertainty and, hence, to facilitate analyzing the images produced from different datasets the array (\ref{ghost0}) is converted into a background-subtracted normalized image structurally identical to $\mathcal{I}'_k$:
\begin{equation}
 \mathcal{I}(x_k,y_k)\equiv\mathcal{I}_k = (\mathcal{I}'_k-\bar{C})/{\bar{A}}.
    \label{ghost}
\end{equation}

In our experiment, quality assessment for the reconstructed ghost images is performed by analyzing the properties of $\mathcal{I}_k$. In accordance with (\ref{Shape}), the spatial resolution of GI is limited by a quantity of $\sim 2\bar{B}\delta_s$.

We apply two measures specifying the image quality. The contrast-to-noise ratio ($CNR$) is a measure traditionally applied to characterize the quality of spatially restricted image parts. In imaging applications, these parts usually refer to transmitting (“In”) or non-transmitting (“Out”) areas of the imaged object. In our experiment, the pixels of these parts are addressed by the indices $k$ belonging to the sets $\{In\}$ or $\{Out\}$,  respectively. The $CNR$ defined as a ratio of the image contrast $K$ to the total error $\sigma$ is found in the form:
\begin{equation}
CNR=\frac{K}{\sigma}=\frac{\langle\mathcal{I}_k\rangle_{\{In\}}-\langle \mathcal{I}_k\rangle_{\{Out\}}} {\sqrt{\sigma_{In}^2+\sigma_{Out}^2}},
\label{CNR}
\end{equation}
where the averaging $\langle\ldots\rangle_{\{In,Out\}}$ is carried out over the sets $\{In\}$ or $\{Out\}$. The quantities $\sigma_{In,Out}^2$ are corresponding variances of $\mathcal{I}_k$. At evaluating $CNR$, we cannot properly assess the influence of strong-noise pixels on the ghost image since they are sparsely dispersed over the SPAD array and the area “In” is essentially small. Therefore, for definiteness, the sets $\{In\}$ and $\{Out\}$ includes only indices of low-noise pixels.

\begin{figure}
    \centering
    \includegraphics[width=0.99\linewidth]{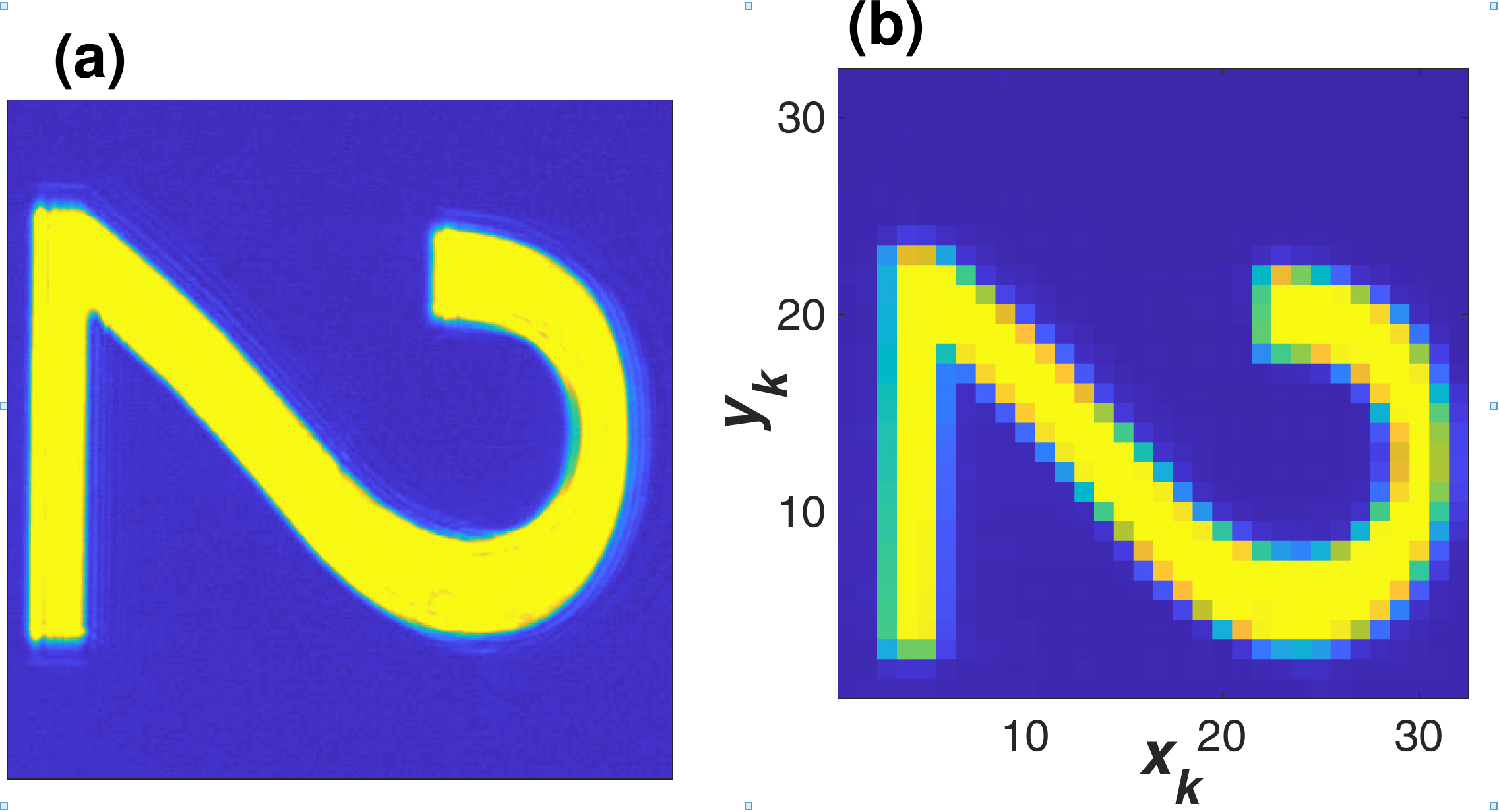}
    \caption{A high-resolution conventional image (a) and a \protect{$32\times32$}-pixel reference image \protect{$I^{(r)}(x_k,y_k)$} (b) of the object.}
    \label{Fig:RefImag}
\end{figure}

The quality of image can also be assessed with help of a correlation coefficient which pixel-by-pixel measures the strength of linear relationship (that is, a structural similarity \cite{wang2004image}) between the analyzed image and a higher-quality reference image of the same object. Unlike $CNR$, 
this measure is applicable to the entire image. We apply the correlation coefficient to evaluate the quality for the entire "ghost" area including highly noised pixels. To produce a reference image of suitable quality, we project a conventional image of the object (back-lighted by a laser beam) with help of a converging lens on the photosensitive matrix of a high-resolution digital camera (UI-3240CP-NIR-GL Rev.2, DS Imaging Development Systems GmbH, pixel size of $5.3$ $\mu m$) at a lens magnification of $\sim1.5$. A high-resolution image (spatial resolution of $\sim 3.5$ $ \mu m$/pixel, this image is illustrated in Fig.~\ref{Fig:RefImag}(a)) stored by the digital camera is converted into a $32\times32$-pixel image $I^{(r)}(x_k,y_k)\equiv I^{(r)}_k$ (resolution of $\sim 45$ $\mu m$/pixel).  The obtained high-contrast and low-noise image $I^{(r)}_k$ (the image is shown in Fig.~\ref{Fig:RefImag}(b)) is then exploited as a reference in order to find the correlation coefficient:  
\begin{equation}
R=\frac{\Bigl|\sum\limits_{k\in \mathit{\{ghost\}}}(I^{(r)}_k-\langle I^{(r)}_k\rangle)(\mathcal{I}_k-\langle \mathcal{I}_k\rangle)\Bigr|}{\sqrt{\sum\limits_{l,k\in \mathit{\{ghost\}}}(I^{(r)}_l-\langle I^{(r)}_l\rangle)^2(\mathcal{I}_k-\langle \mathcal{I}_k\rangle)^2}},
    \label{CorrCoeff}
\end{equation}
where the summation is done over the "ghost" part of the SPAD area. In equation (\ref{CorrCoeff}) brackets $\langle \ldots\rangle$ denote averaging over the "ghost" area pixels. To achieve an accurate correspondence between the ghost and reference images, (i.e., to ensure the same position and the same scaling factor for the both images on the $32\times32$ pixel grid) the correlation coefficient $R$ is maximized at image converting. The error of $R$ evaluation  is less than $0.01$. 

Overall the following procedure of ghost-image reconstruction is implemented. First of all, a dataset  is prepared. The prepared dataset is applied to calculate $G^{(1)}(m)$ and $G^{(2)}(m,k)$ and find normalized correlation function $g^{(2)}(m,k)$ as described in equations (\ref{G12}) and (\ref{g2}). After that the averaged parameters $\bar{A}$, $\bar{B}$, $\bar{C}$ and $\overline{A/C}$ of correlation function \eqref{Shape} are evaluated. Then, a ghost image $\mathcal{I}_k$ is computed in accordance with equations (\ref{ghost0}) and (\ref{ghost}). Finally, quality of the obtained image is examined visually and estimated using  $CNR$ (\ref{CNR}) and correlation coefficient (\ref{CorrCoeff}). 

\section{Results and discussion}
In our GI experiment, the light beams are registered under conditions where $\langle n_f \rangle\approx8$. The registered data stream $\mathcal{F}$ is collected over 186 minutes, stored on the computer and later post-processed. At post-processing, this dataset is used to prepare an unfiltered dataset $F_{0}$ and then to produce a number of frame-filtered sets $F(n^*)$.  

\begin{figure*}
    \centering
        \begin{subfigure}{0.3\textwidth}
    \includegraphics[width=\linewidth]{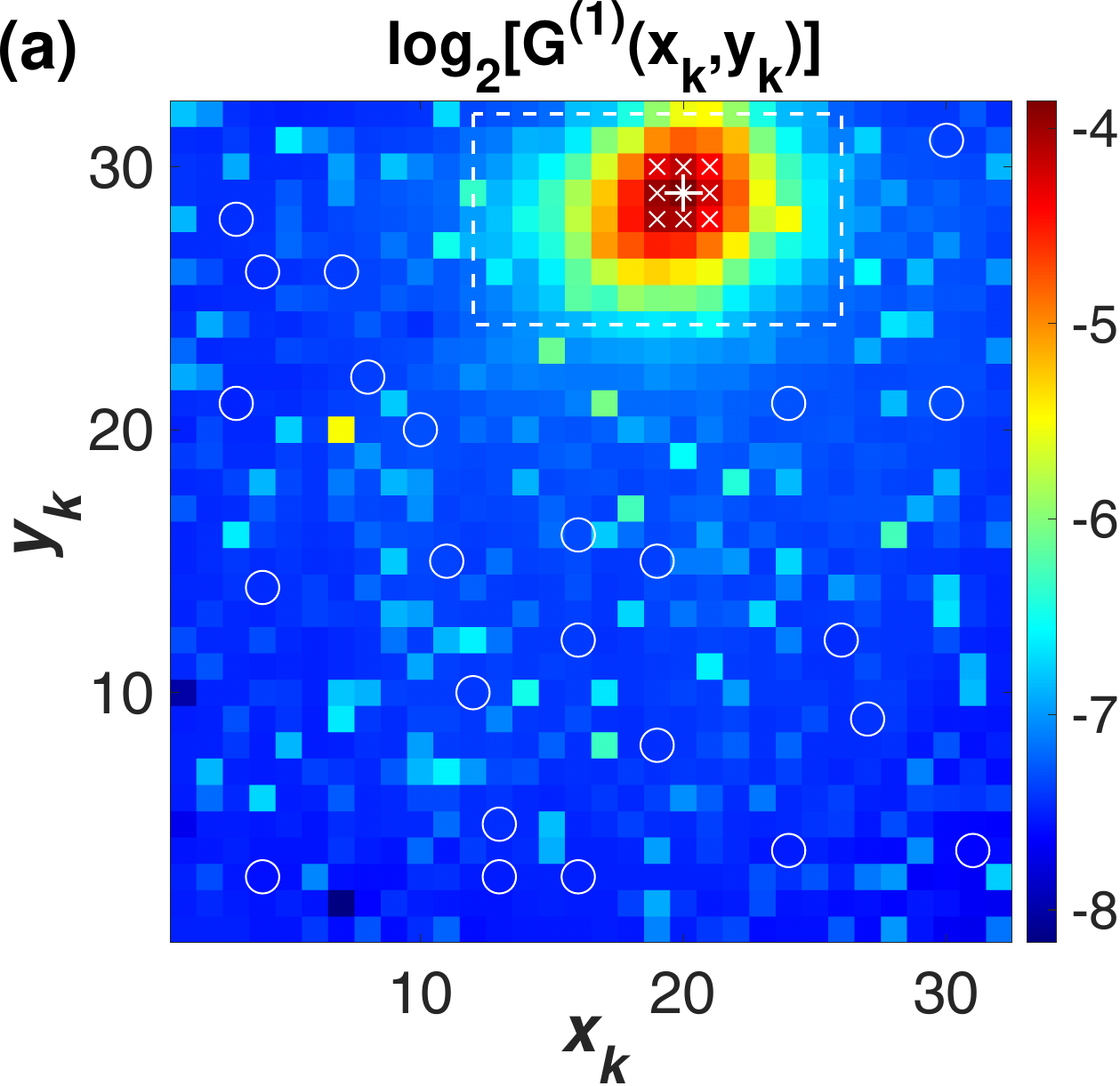}
    \end{subfigure}
        \begin{subfigure}{0.3\textwidth}
    \includegraphics[width=\linewidth]{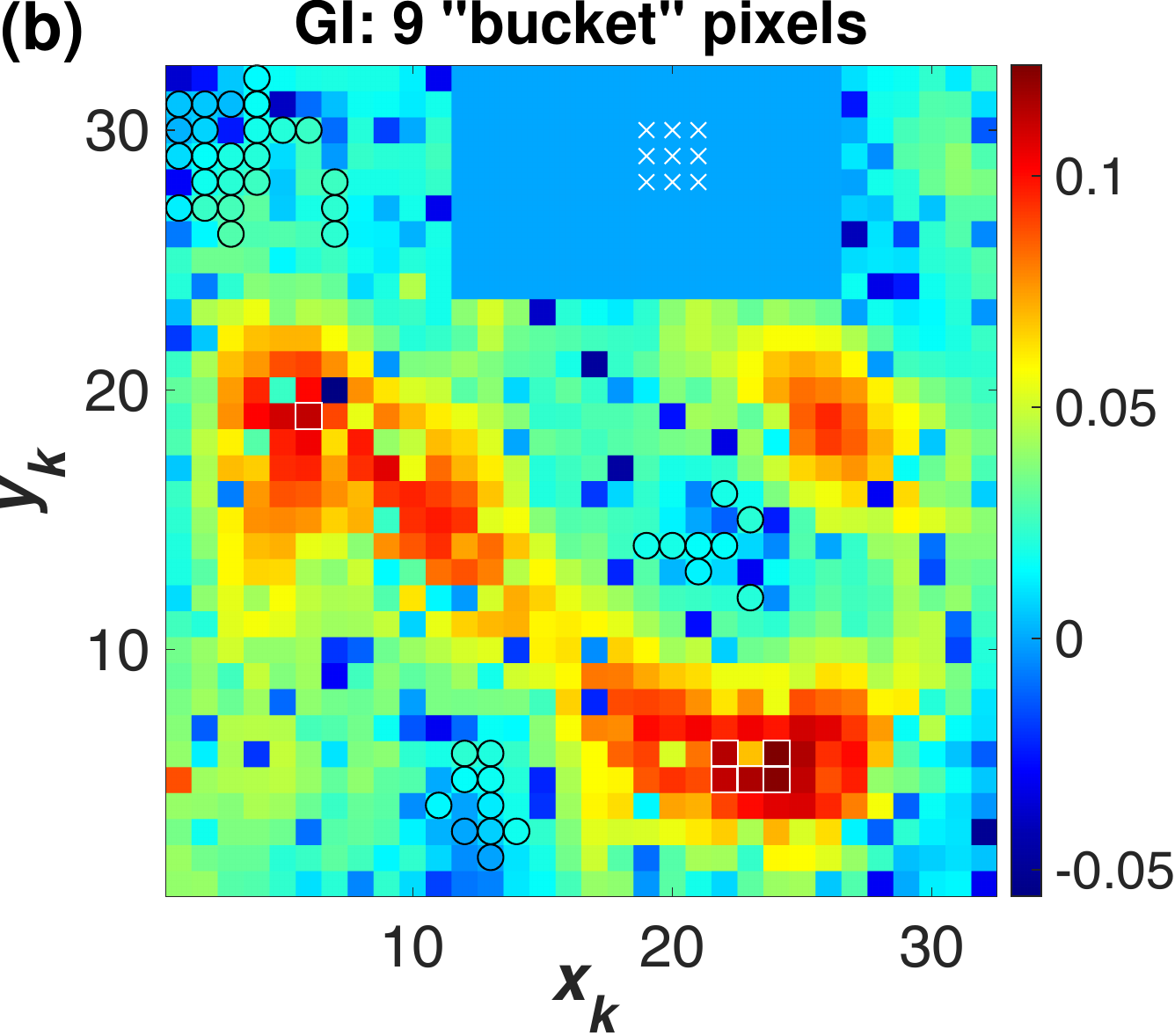}
    \end{subfigure}
       \begin{subfigure}{0.3\textwidth}
    \includegraphics[width=\linewidth]{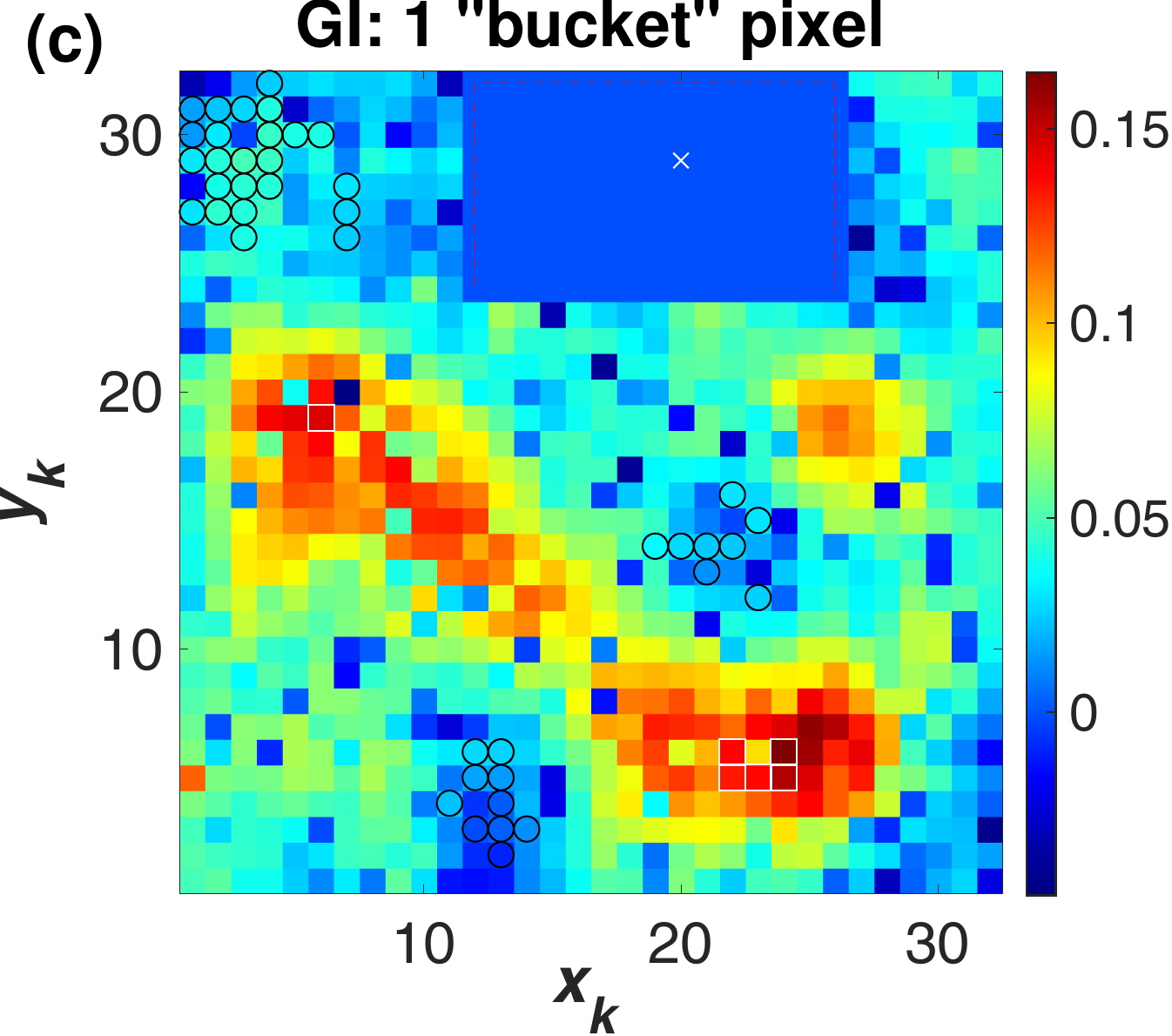}
    \end{subfigure}        
    
   \caption{{GI based on the unfiltered dataset $F_{0}$. The white x’s show the pixels belonging to the  "bucket” region. The white squares and black circles show, respectively, the pixels belonging to the sets $\{In\}$ and $\{Out\}$. \textbf{Panel (a)}: distribution of \protect{ $\log_2[G^{(1)}(x_k,y_k)]$}; the white cross shows the point (20,29) corresponding to the maximal intensity of the object-beam spot;  the white dashed line indicates a rectangle containing the “flare” region; the white circles show the “probing” pixels. \textbf{Panel (b)}: ghost image \protect{$\mathcal{I}(x_k,y_k)$}; white crosses indicate positions of 9 pixels chosen as a "bucket". \textbf{Panel (c)}: ghost image reconstructed with the single-pixel "bucket" marked by x in the "flare" area.}}
   \label{fig:fig3}
\end{figure*}
The maximum count rate for the object-beam spot on the SPAD array is realized near the point $(20,29)$ (here and further we denote the pixels coordinates $x$ and $y$ just by a pair of indices $(x,y)$). The "flare" area is taken as a rectangle with vertices at the pixels (12,32), (12,24), (26,24), (26,32). The rectangle contains $15\times9=135$ pixels. So, the remaining "ghost" area includes $J_{max} =1024 - 135 =889$ pixels. Twenty four low-noise pixels uniformly and randomly distributed across the "ghost" area are selected as the "probing" pixels $m\in \{m^{*}\}$ required to determine the averaged parameters $\bar{A}$, $\bar{B}$, $\bar{C}$ and $\overline{A/C}$.

In the study, we apply two sets $\{B\}$ of "bucket" pixels. Eight low-noise pixels around the pixel $(20,29)$ and this pixel are used as a "bucket" set. On the other hand, to highlight advantages offered by our approach, we also demonstrate a traditional GI technique. For that we just choose as a "bucket" detector a single pixel in the "bucket" region. This pixel is the point $(20,29)$. The "flare" area, the exploited "probing" and "bucket" pixels are indicated in Fig~\ref{fig:fig3}(a). 

\subsection{GI with unfiltered data}
 
 The total number of frames in the unfiltered dataset $F_{0}$ is $N_{0}\sim1.9\times 10^9$. The per-frame probabilities $G^{(1)}(x_k,y_k)\equiv G^{(1)}(k)$ obtained for all pixels of the SPAD array from this dataset are shown in Fig.~\ref{fig:fig3}(a). The averaged values of shape parameters and the corresponding standard deviations are as follows:  $\bar{A} \approx 0.780$, $\sigma_A \approx 0.047$;   $\bar{B}\approx 1.72$, $\sigma _B\approx 0.085$; and   $\bar{C}\approx 1.085$, $\sigma_C\approx 0.0046$. According to the obtained data, the spatial resolution of the image computed from the data $F_{0}$ is of $\sim 2\bar{B}\delta_s\approx 3.44 \delta_s \approx 0.15$ mm. The average per-frame probability of counts detected by the "bucket" pixels $\langle G^{(1)}(k)\rangle_{k\in\{B\}}\approx 0.057$ (this quantity is obtained by averaging of $\langle G^{(1)}(k)\rangle$ over the nine “bucket” pixels) is $\sim 300$ times higher than the corresponding averaged probability of dark counts $\langle G_{DC}^{(1)}(k)\rangle_{k\in\{B\}}\approx 1.8 \times 10^{-4}$.  
 
 The computed ghost image $\mathcal{I}(x_k,y_k)$ is represented in Fig.~\ref{fig:fig3}(b). The visual examination of this figure shows that the imaged object is recognizable in the ghost image. However, the presented figure clearly demonstrates a serious drawback. The number of noisy pixels is so high that they have a strong destructive effect on the quality of the entire image.  According to Appendix, the efficient number of pixels distinguishable (due to the dark-count effect) in the ghost image reconstructed from the dataset $F_{0}$ is of $\sim 100$. In addition, the spatial resolution is clearly suffering.  The resolution is not high enough to resolve all parts of the object. The "tail" of the imaged digit "2" 
 (a $\sim 0.1$ mm wide strip) is not visible in the reconstructed image. 
 
The quality of the obtained image $\mathcal{I}_k$ is estimated numerically in terms of $CNR$ (\ref{CNR}) and structure similarity with the reference image (\ref{CorrCoeff}). In the $CNR$ evaluation we exploit 6 low-noise pixels for the set $\{In\}$ and 43 low-noise pixels for the set $\{Out\}$ (these pixels are indicated in Fig.~\ref{fig:fig3}(b)). The $\{In\}$  pixels are located near the maxima of the observed ghost image. For these pixels sets we have $\langle\mathcal{I}_k\rangle_{\{In\}}\approx 0.1170$, $\sigma_{
In}\approx 0.0046$ and  $\langle\mathcal{I}_k\rangle_{\{Out\}}\approx 0.0163$, $\sigma_{Out}\approx 0.0073$. According to (\ref{CNR}), the image $CNR\approx 11.7$ is achieved at the contrast of $K\approx 0.101$ and the total error $\sigma \approx 0.0086$. Along our estimation (\ref{CorrCoeff}), the coefficient of correlation $R$ between the image $\mathcal{I}_k$ and the reference image $\mathcal{I}_k^{(r)}$ is of $\sim 0.62$. 

\begin{figure}
    \centering
    \includegraphics[width=0.99\linewidth]{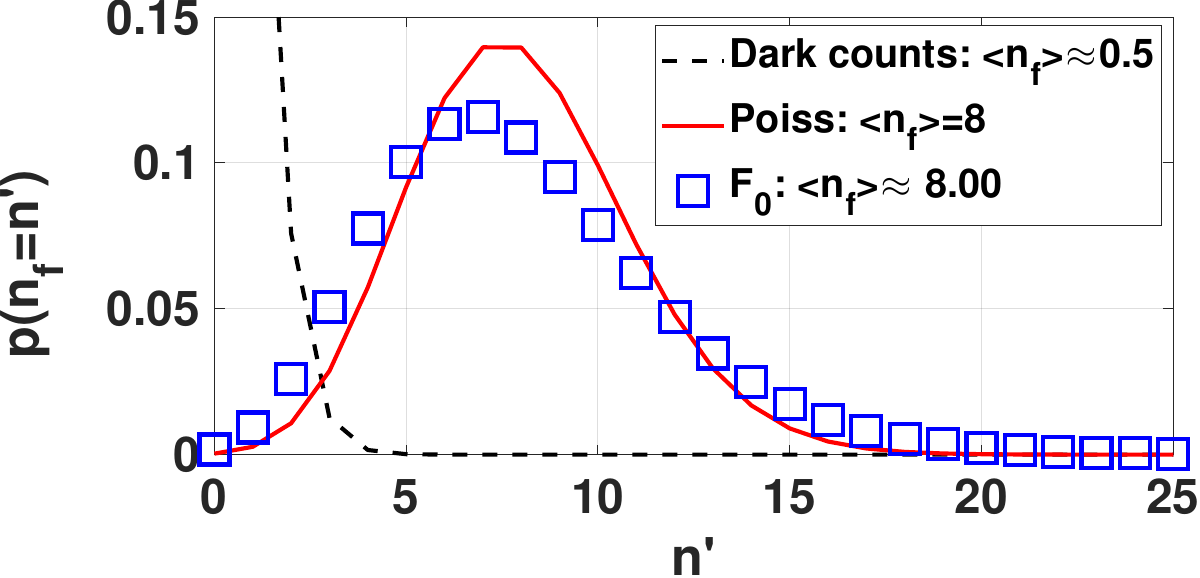}
    \caption{Distribution of per-frame probabilities $p(n_f=n')$ over realizations of $n_f$ (blue squares) for the unfiltered dataset $F_{0}$. The Poisson probability distribution of the same $\langle n_f\rangle$ are shown by the red solid line. The Poisson distribution that represents the dark counts typical for the applied SPAD array is depicted by the black dashed line}
    \label{fig:StatAnal}
\end{figure}

The unfiltered data $F_{0}$ is a composition of frames with different values of $n_f$.  Fig.~\ref{fig:StatAnal} shows the probability distribution that expresses a per-frame probability \eqref{ProbDistr}. The shape of this distribution is rather close to the shape of the Poisson distribution with the same average number of counts per frame. The obtained $p(n_f=n')$ distribution has a maximum near $n' =7$ and a half-maximum width of $\sim 9$ . 

\subsection{GI with filtered data}

In this subsection, we exploit frame-filtered datasets $F(n^*)$ that are produced by filtering the data $F_{0}$ considered in the previous subsection.  For the exploited data, the parameter $n^*$ ranges from 2 to 24 (see  Fig.\ref{fig:StatAnal}). 

\begin{figure}
    \centering
    \includegraphics[width=0.99\linewidth]{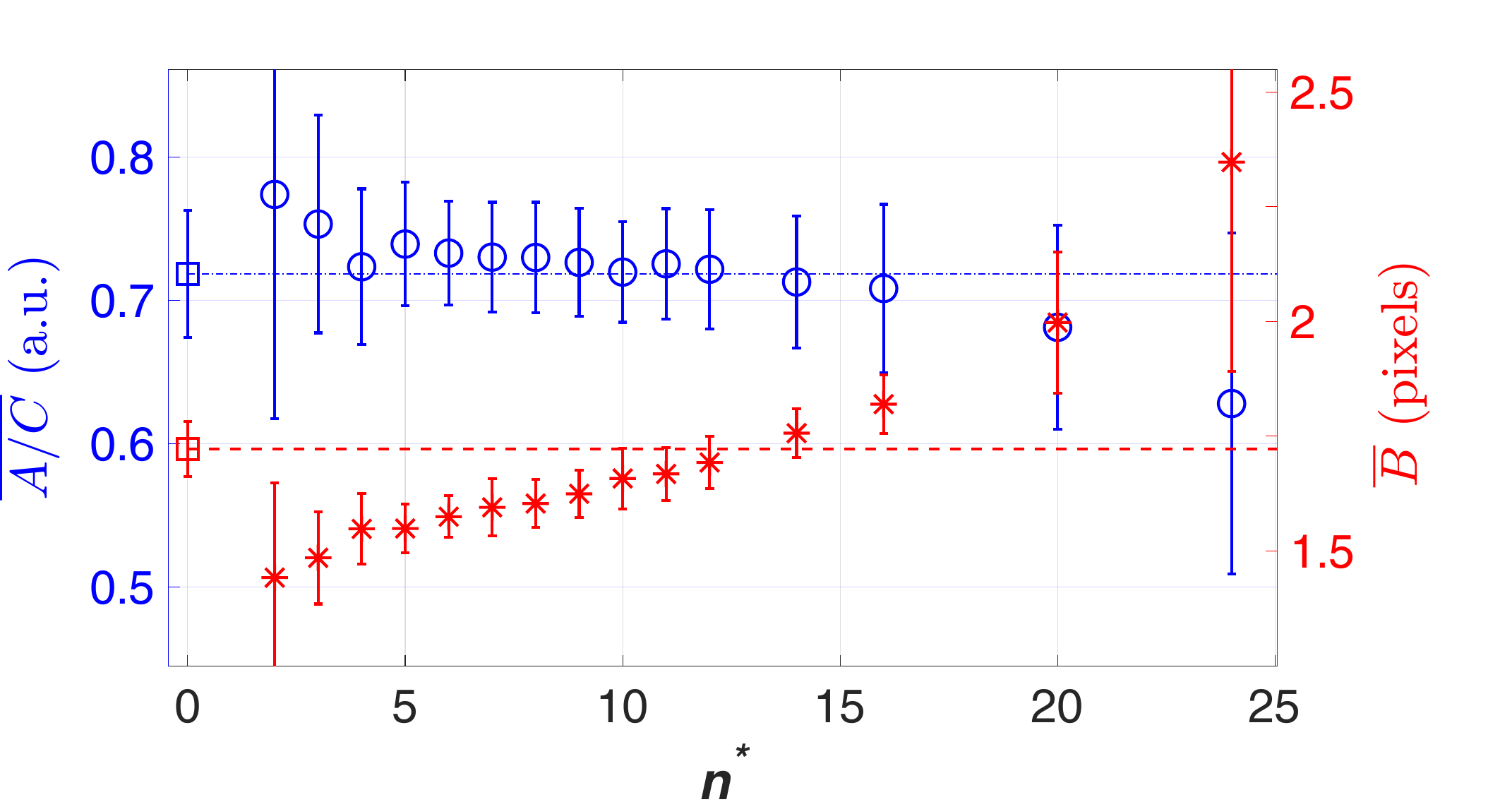}
    \caption{
Dependence on $n^*$ for averaged ratio $\overline{A/C}$ ( left Y axis: blue circles) and half-width $\bar{B}$ ( right Y axis: red stars) obtained from frame-filtered datasets $F(n^*)$.  The relevant parameters acquired from the unfiltered data $F_{0}$ are shown by a blue dash-dotted line with a blue square (left Y axis: $\overline{A/C}$) and a red dashed line with a red square (right Y axis: $\bar{B}$).}
    \label{fig:Shape}
\end{figure}

The datasets $F(n^*)$ are used to clarify how the dataset composition may affect the correlation function \eqref{Shape}.  Fig.\ref{fig:Shape} shows the dependence of the half-width $\bar{B}$ and the ratio $\overline{A/C}$  on $n^*$ (the scatters show the corresponding root-mean-squared deviations). As can be seen from this figure, the parametric ratio $\overline{A/C}$, within the error of its determination, does not depend on $n^*$. 
On the other hand, the half-width $\bar{B}$ monotonically increases with $n^*$ from 1.4 to 2 pixels. As a result, the spatial resolution of GI will worsen from $\sim 0.12$ mm up to $\sim0.18$ mm with the parameter $n^*$. The GI based on a frame-filtered dataset with $n^*<12$ must exhibit better spatial resolution than the GI reconstructed from the unfiltered data $F_{0}$. The observed dispersion-like phenomenon for the half-width $\bar{B}$  may be associated with specific optical properties of the rotating disk D. 
\begin{figure}
    \centering
    \includegraphics[width=0.99\linewidth]{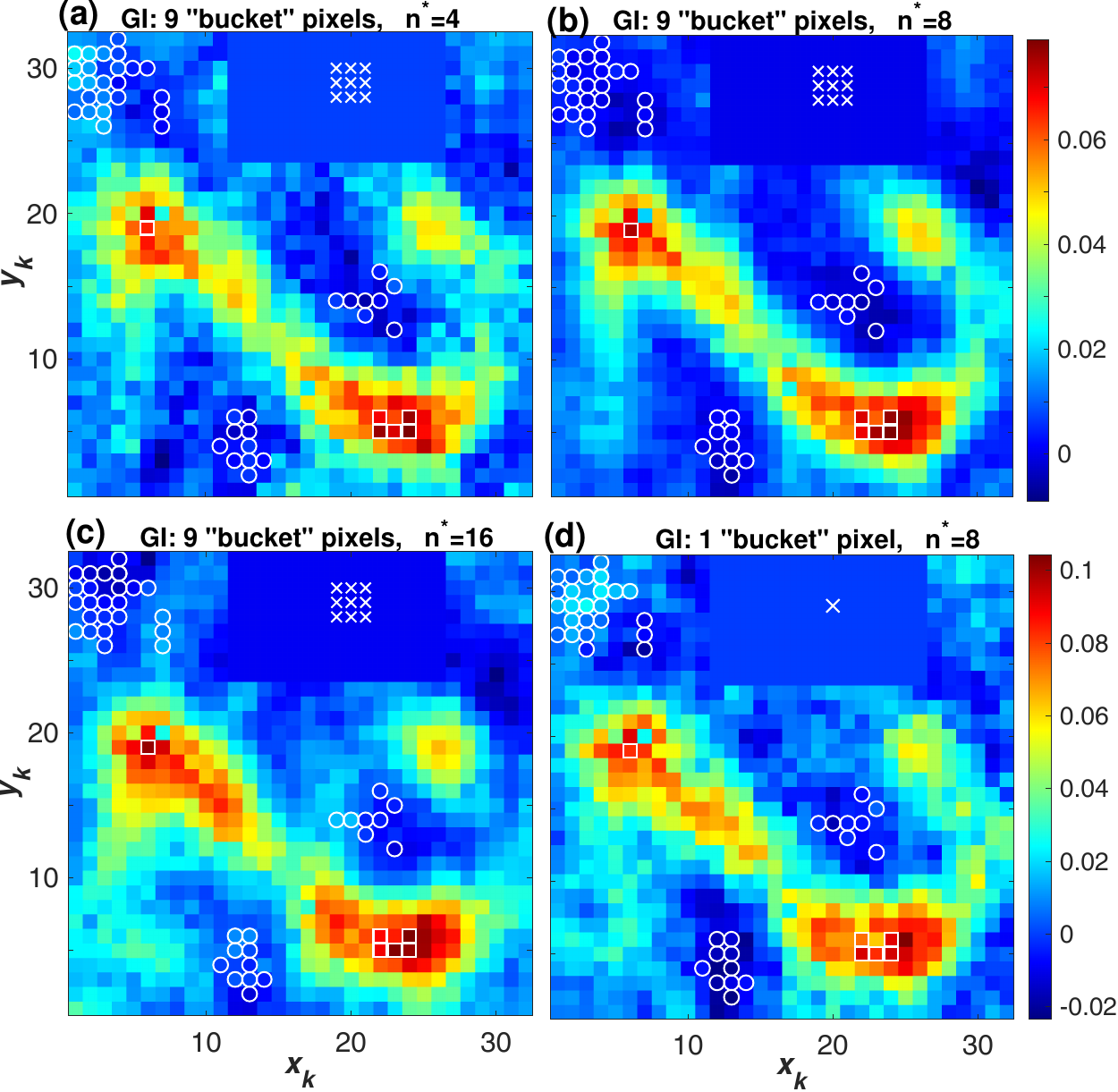}
    \caption{Ghost images reconstructed from the frame-filtered datasets $F(4)$ (a), $F(8)$ (b) and $F(16)$ (c).  Ghost image computed from the frame-filtered dataset $F(8)$ with the single-pixel "bucket" (d). The white squares, circles and x's show, respectively, the pixels belonging to the sets $\{In\}$, $\{Out\}$ and $\{B\}$.}
    \label{fig:fig5}
\end{figure}
The datasets $F(n^*)$ are used to reconstruct ghost images. Fig.~\ref{fig:fig5} demonstrates images $\mathcal{I}_k$ computed from frame-filtered datasets $F(n^*)$ at $n^*=4$, $8$ and $16$. In terms of contrast, spatial resolution, and number of distinguishable noised pixels, the presented images look significantly better compared to the ghost image obtained from unfiltered data. Image qualities (such as spatial resolution and contrast) depend on the parameter $n^*$. Based on visual inspection of the presented images one can conclude that ghost images of the highest quality are realized for datasets $F(n^*)$ with $n^*\sim 8$. As can be seen from Fig~\ref{fig:fig5} (b), the spatial resolution of the image $\mathcal{I}_k$ based on the dataset F(8) is high enough to resolve all parts of the imaged object. In accordance with Appendix, only a few pixels of the image reconstructed from F(8) can be identified as pixels affected by dark counts.

\begin{figure}
    \centering
    \includegraphics[width=0.99\linewidth]{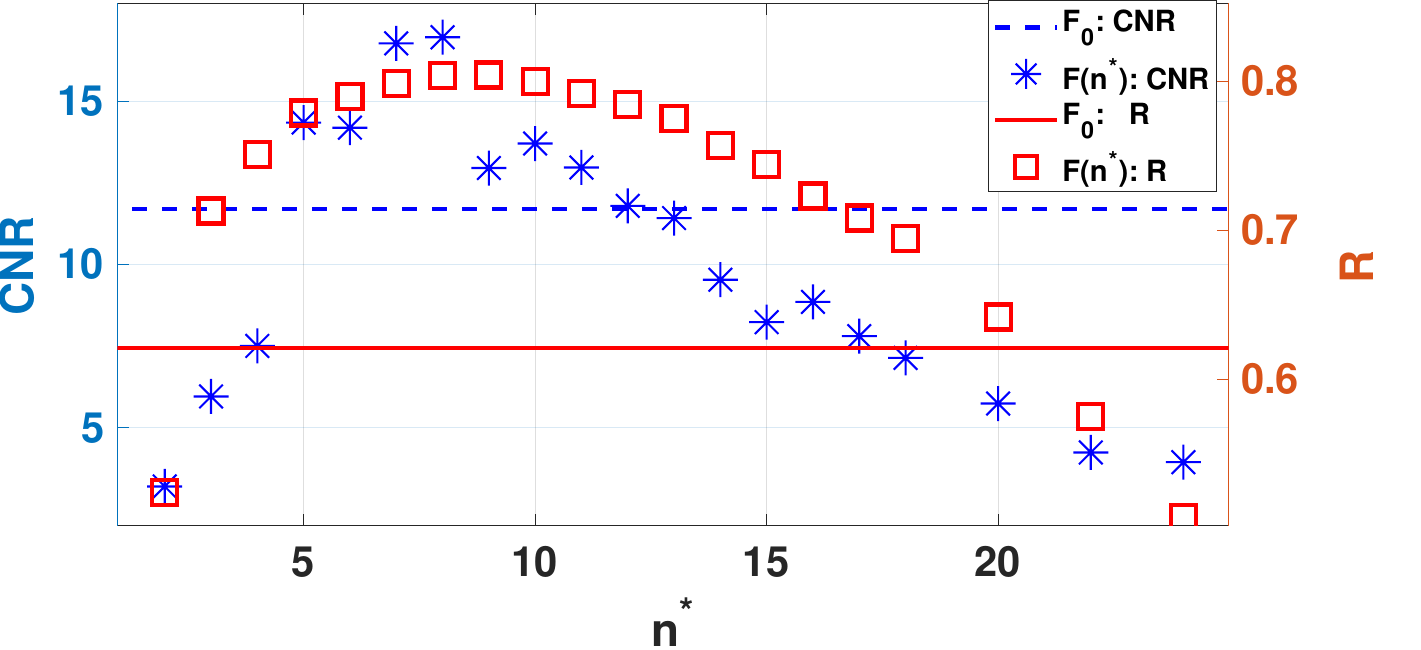}
    \caption{$CNR$ (left Y axis: blue stars) and correlation coefficient $R$ (right Y axis: red squares) as functions of $n^*$ for ghost images reconstructed from frame-filtered datasets $F(n^*)$. The blue dashed and red solid lines indicates $CNR$ (left Y axis) and $R$ (right Y axis) for the unfiltered data $F_{0}$ . }
    \label{fig:fig6}
\end{figure}

This conclusion is corroborated with the results of numeric evaluation of $CNR$ (\ref{CNR}) and the correlation coefficient (\ref{CorrCoeff}) represented in Fig.~\ref{fig:fig6}. In the $CNR$ evaluation, the pixels $k\in \{In\}$ and $k\in \{Out\}$  are the same pixels as in the previous subsection. Generally, for the ghost image based on the frame-filtered data, the maximum $CNR$ and correlation coefficient $R$  are achieved near those values of $n^*$ that correspond to the maximum probability $p(n_f=n')$ of the distribution analyzed in the previous subsection and represented in Fig.\ref{fig:StatAnal}. As shown in Fig.~\ref{fig:fig6} (left Y axis), compared to the "unfiltered" image, the "frame-filtered" image $\mathcal{I}_k$ has a higher $CNR$ when the parameter $n^*$ ranges from 5 to 10. At $n^*=8$, the filtered GI reaches its maximal $CNR\approx 17.0$ ($\langle\mathcal{I}_k\rangle_{\{In\}}\approx 0.0744$, $\sigma_{
In}\approx 0.0033$, $\langle\mathcal{I}_k\rangle_{\{Out\}}\approx 0.0009$, $\sigma_{Out}\approx 0.0028$, $K\approx 0.0735$, $\sigma\approx 0.0043$) which is nearly $45\%$ higher as compared to the $CNR$ of unfiltered GI. 

Unlike $CNR$, the measure of structural similarity  (\ref{CorrCoeff}) takes into account the effect of strong-noise pixels on the image. Therefore, the range $3\leq n^* \leq 19$, where the "low-noise" filtered GI leads to a higher coefficient R (as compared to the "noisy" unfiltered GI), covers almost all of the considered values of $n^*$.  Like $CNR$, the structural similarity between the reference and ghost images attains its maximum at $n^*=8$ leading to $R\approx 0.8$ (right Y axis of Fig.~\ref{fig:fig6}).  

\subsection{GI with a single-pixel "bucket"}

The per-frame probability of counts detected by the single "bucket" pixel  $G^{(1)}(20,29)\approx 0.069$ obtained from the unfiltered data $F_0$ is substantially high compared to the corresponding probability of dark counts $ G_{DC}^{(1)}(20,29)\approx 4.5 \times 10^{-5}$.  The ghost images computed with the single-pixel "bucket" from the unfiltered data $F_{0}$ and the filtered dataset $F(8)$ are shown in Fig ~\ref{fig:fig3}(c) and Fig ~\ref{fig:fig5}(d), respectively. The resulting ghost image obtained with the data $F_{0}$  is the image that one gets with traditional ghost imaging technique for the case. 
It seems close to the GI of Fig ~\ref{fig:fig3}(b) obtained with our  approach  for the data $F_{0}$. Indeed, the correlation coefficient is only slightly lower, $R\approx 0.6$ in comparison to the correlation for the  multiple-pixel "bucket" image ( $R\approx 0.62$). However, $CNR$ is nearly twice lower, $CNR\approx 6.73$. 

Using the filtered dataset $F(8)$ improves the correlation giving $R\approx 0.724$. However, simultaneously it leads to degradation of the contrast, $CNR\approx 5.46$. Thus, a single-pixel "bucket" imaging modality for the case cannot ensure the quality of image given by averaging over several GI  in our approach. 

\section{Conclusions}
We have presented an approach to ghost imaging and analyzed the approach capabilities in an experiment where the light intensity is so low that the dark counts have a noticeable effect on the imaging. 
We have demonstrated that filtering the acquired data with respect to the number of counts per frame and obtaining the image as a weighted average over the images for different "bucket" pixels, one can significantly reduce the  destructive effect of dark counts on the ghost image, improve the image contrast, spatial resolution and image similarity to the reference image. 
The proposed approach allowed us inferring ghost images for dark counts  reaching $20\%$ of the the total "bucket" counts and with the photon fluxes of just few photons per time-frame.  In our opinion, using multiple bucket detectors instead of just one is compatible with most applications of ghost imaging and will allow one to greatly enhance imaging quality in the extremely-low-light condition. These results can be extended to computational \cite{Erkmen:10,PhysRevA.78.061802,2019NaPho..13...13E,Bruschini} and temporal \cite{2016NaPho..10..167R,Devaux:16,Horoshko:23}  ghost imaging.
Our findings are important for developing correlation imaging schemes with light on a-few-photon level, first of all, for quantum sensing applications. 

We feel that as a concluding remark it would be useful to emphasize our motivation. As it is shown in  Appendix, the proposed
frame-filtering approach improves considerably the reconstructed ghost
images and, simultaneously, has no effect on the quality of $G^{(1)}$. Thus, we do not expect any improvement compared to the conventional imaging with the same light source and detection system \cite{Moreau:18}. Our motivation was to improve the quality of the ghost imaging with very low photon fluxes using very realistic noisy SPADs with inhomogeneous quality of pixels. We see this task as quite actual and practical for improving ghost imaging in its application niche.

\section*{Funding}
V.S., S.K., V.C. and A. S. acknowledge financial support from the State Research Program "Convergence" subprogram 3.01.2.  D.M. acknowledges financial support from {the BRFFR project F23UZB-064. This work was supported by the European Commission through the SUPERTWIN Project under Grant 686731.} 

\section*{Disclosures}
The authors declare no conflicts of interest.

\section*{Data availability}
Data underlying the results presented in this paper are not publicly available at this time but may be obtained from the authors upon reasonable request.

\appendix

\section{Dark counts of the SPAD array and their effect on reconstructed ghost images}
\begin{figure}
    \centering
    \includegraphics[width=0.75\linewidth]{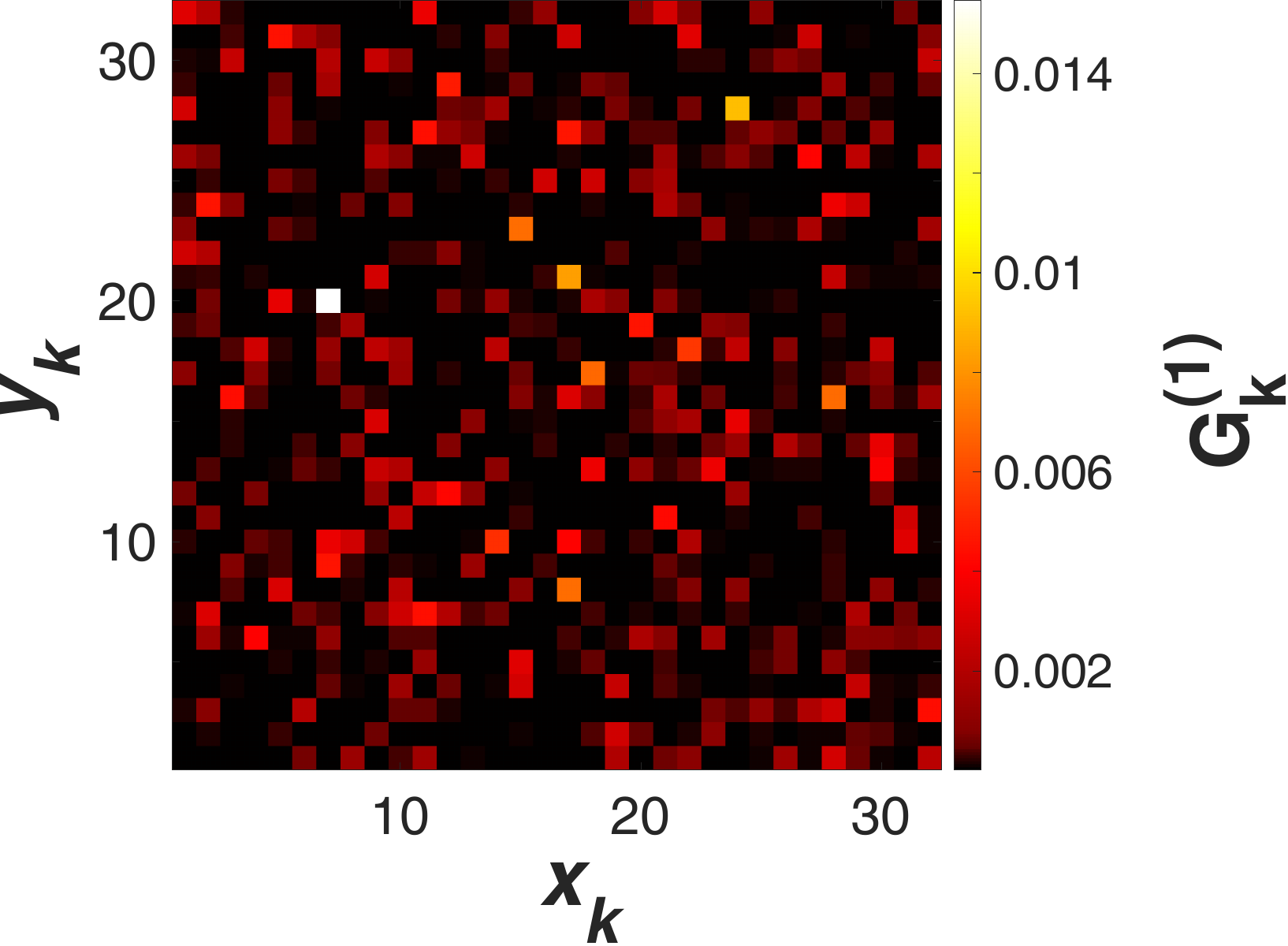}
    \caption{Per-frame dark counts distribution $G^{(1)}$ over the SPAD pixels. The colorbar shows the corresponding normalized values.}
    \label{fig:DC_G1}
\end{figure}

To analyze the effect of dark-counts (DC) for the SPAD array we, first of all, measure the probability of counts for each SPAD pixel under conditions where the detection events are DC. To ensure this condition, the SPAD array is blocked by a light-tight screen from any light source (including ambient light). The registered data are stored in the computer as a frame-unfiltered dataset $F_{DC}$ including $\sim 2.8\times 10^{9}$ frames. The per-frame probabilities of DC $G^{(1)}_{DC}(k)$ for all SPAD pixels are found from the dataset $F_{DC}$, as described in Section III. The distribution of measured $G^{(1)}_{DC}(x_k,y_k)$ over the pixels of the SPAD array is shown in Fig~\ref{fig:DC_G1}. The bright pixels of this distribution are strongly noised pixels of the SPAD array. The brightest pixel is the pixel $k=212$ with coordinates $x_k=7, y_k=20$. One can see in Fig~\ref{fig:DC_G1} that distribution of DC rates over the SPAD pixels is quite inhomogeneous and randomized. The measured probabilities $G^{(1)}_{DC}(k)$ are in a wide range of values from $\sim 10^{-5}$ to  $\sim 10^{-2}.$ The average DC probability per pixel and per frame reaches a value of $\langle G^{(1)}_{DC}\rangle \approx 4.9 \times 10^{-4}$ (or approximately 0.5 per frame). The dependence of per-frame probability $p(n_f=n')$ on $n'$ for this data is represented by the Poisson distribution with $\langle n_f\rangle \approx 0.5$ and shown in Fig.\ref{fig:StatAnal}.

\begin{figure}
    \centering
    \includegraphics[width=0.99\linewidth]{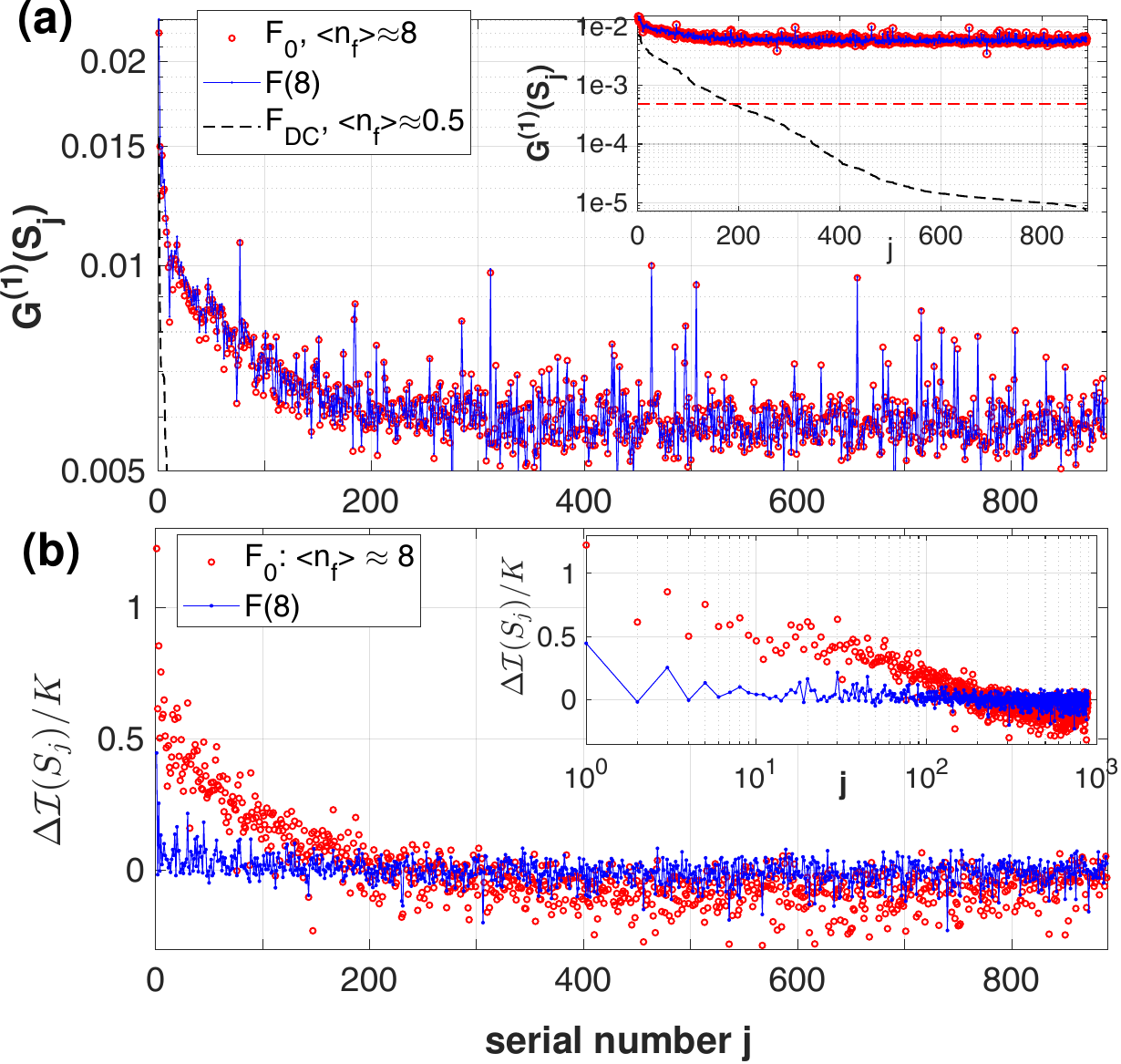}
    \caption{“DC-ordered” sequences $G_{DC}^{(1)}(S_j)$ (a, black dashed line), $G^{(1)}(S_j)$ (a) and $\Delta \mathcal{I}(S_j)$ (b) as functions of the serial number $j$. The red circles show the sequences calculated from the unfiltered data $F_{0}$. The blue solid line and dots indicate the sequences found from the frame-filtered dataset $F(8)$. The inset of panel (a) demonstrates the same data (as in the main panel) at a larger Y-axis scale. The red dashed line in this inset indicates the average probability $\langle G^{(1)}_{DC}\rangle$. The inset of panel (b) shows the same data (as in the main panel) in a log scale on the X-axis. }
    \label{fig:DCOrder}
\end{figure}

In general, despite the substantial inhomogeneity and randomness, the influence of DC on the measurement can be evaluated if the evaluation operates with specially organized sequences of pixels. In these sequences, the pixels are arranged so that their dark count rates $G^{(1)}_{DC}(k)$ are ordered in descending order (“DC-ordering”): 
\begin{equation}
G^{(1)}_{DC}(k) \xrightarrow{DC-ordering}G^{(1)}_{DC}(S_j),
\end{equation}
where $j$ is the serial number for each term of the sequence $G^{(1)}_{DC}(S_j)$. The sort index $S_j$ specifies how the data $G^{(1)}_{DC}(k)$ should be rearranged to obtain the ordered sequence $G^{(1)}_{DC}(S_j)$. The first terms of the DC-ordered sequence belong to strongly noisy pixels, the last to low-noise pixels. 

The DC effect on a measured quantity $Q(k)$ is evaluated by analyzing a “DC-ordered” sequence $Q(S_j)$. For the pixels with a weak influence of DC on $Q$ (e.g., for low-noise pixels with $j\gg 1$) the measured $Q(S_j)$ fluctuates around a certain value $Q_0$, independent of $j$. On the other hand, for a pixel $k^{\prime}$ where the influence of DC on $Q$ is significant, the quantity $Q(k^{\prime})$ is a function of $G^{(1)}_{DC}(k^{\prime})$. For several pixels with strong DC effect, the DC-ordered sequence $Q(S_j)$  explicitly reproduces the sequence $G^{(1)}_{DC}(S_j)$ within a $j$ index range $\nu_Q$. In a sense, the deviation $\chi_Q=max(|Q(S_j)-Q_0|)$ realized near $j=1$ and the width of the range $\nu_Q$ can be accepted as measures of the DC effect. The width $\nu_Q$ can be associated, for instance, with an efficient number of noised pixels that have a destructive effect on the measured quantity $Q$. The deviation $\chi_Q$ characterizes the effect strength for the "hot" pixels. 

To evaluate the DC effect on a reconstructed ghost image $\mathcal{I}(x_k,y_k)$ we compute a pixel-by-pixel image difference $\Delta \mathcal{I}(k)\equiv\Delta \mathcal{I}(x_k,y_k)=\tilde{\mathcal{I}}(x_k,y_k)-\mathcal{I}(x_k,y_k)$. Here, $\tilde{\mathcal{I}}(x_k,y_k)$ is an image obtained from $\mathcal{I}(x_k,y_k)$ by a 2D smoothing in which the convolution kernel is a 3$\times$3 pixel square. The quantity $\Delta \mathcal{I}(k)$ is insensitive to pattern variation for the imaged object and approximately equal to fluctuation in $\mathcal{I}(x_k,y_k)$ between the neighboring pixels. Owing to these properties, the difference $\Delta \mathcal{I}$ is well adapted to register one-pixel-ranged correlation drops typical for strongly noised pixels sparsely dispersed across the SPAD array. We analyze the dependence of the “DC-ordered” difference $\Delta \mathcal{I}(S_j)$ on the index $j$. In a similar way, the DC influence on the measured probabilities $G^{(1)}$ can be evaluated by considering the sequence $G^{(1)}(S_j)$ as a function of $j$. The “DC-ordered” sequences $G^{(1)}(S_j)$ and $\Delta \mathcal{I}(S_j)$ calculated for the pixels $k\in \{ghost\}$ ($j=1,\ldots,J_{max}$) from frame-filtered and unfiltered datasets (they are datasets $F(8)$ and $F_{0}$ applied for ghost image computation in Section IV) are represented in Fig.~\ref{fig:DCOrder}.  For illustrative purposes (to demonstrate the effect manifestation in comparison with the image contrast $K$), the difference $\Delta \mathcal{I}(S_j)$ in the figure is normalized by $K$.

Regardless of the type of filtering (with or without frame filtering), the DC have an equal influence on the probabilities $G^{(1)}$ obtained from data with the same average numbers $\langle n_f\rangle$. As shown in Fig.~\ref{fig:DCOrder} (a), the dependencies of $G^{(1)}(S_j)$ on the $j$ index for filtered and unfiltered data ($F(8)$ and $F_{0}$ with $\langle n_f\rangle\approx8$) are very similar. The efficient number of pixels distinguishable due to the DC effect in the imaged distribution $G^{(1)}(x_k,y_k)$ is $\nu_{G_1} \sim 100$ for the both datasets. Contrary to $G^{(1)}$, the manifestation of DC effect in ghost images (see Fig.~\ref{fig:DCOrder} (b)) depends significantly on what data (frame-filtered or not) is used to compute the images. The efficient number of pixels distinguishable (due to the DC effect) in the ghost image computed from the unfiltered data $F_{0}$ is $\nu_{\Delta \mathcal{I}} \sim \nu_{G_1} \sim 100$. 
For the frame-filtered dataset $F(8)$, the manifestation of the DC effect is essentially small. In fact, only a few pixels can be identified that are affected by DC. The deviation $\chi_{\Delta \mathcal{I}}$ realized for the "hottest" pixel $k = 212$ is more than three times smaller than for unfiltered data. Such a dramatic reducing (compared to $F_{0}$) in the DC effect is observed for all frame-filtered datasets used in Section IV at GI reconstructing. All pixels with the index $j>300$ are accepted to be low-noise.


\bibliography{ghbib}

\begin{thebibliography}{37}%
\makeatletter
\providecommand \@ifxundefined [1]{%
 \@ifx{#1\undefined}
}%
\providecommand \@ifnum [1]{%
 \ifnum #1\expandafter \@firstoftwo
 \else \expandafter \@secondoftwo
 \fi
}%
\providecommand \@ifx [1]{%
 \ifx #1\expandafter \@firstoftwo
 \else \expandafter \@secondoftwo
 \fi
}%
\providecommand \natexlab [1]{#1}%
\providecommand \enquote  [1]{``#1''}%
\providecommand \bibnamefont  [1]{#1}%
\providecommand \bibfnamefont [1]{#1}%
\providecommand \citenamefont [1]{#1}%
\providecommand \href@noop [0]{\@secondoftwo}%
\providecommand \href [0]{\begingroup \@sanitize@url \@href}%
\providecommand \@href[1]{\@@startlink{#1}\@@href}%
\providecommand \@@href[1]{\endgroup#1\@@endlink}%
\providecommand \@sanitize@url [0]{\catcode `\\12\catcode `\$12\catcode `\&12\catcode `\#12\catcode `\^12\catcode `\_12\catcode `\%12\relax}%
\providecommand \@@startlink[1]{}%
\providecommand \@@endlink[0]{}%
\providecommand \url  [0]{\begingroup\@sanitize@url \@url }%
\providecommand \@url [1]{\endgroup\@href {#1}{\urlprefix }}%
\providecommand \urlprefix  [0]{URL }%
\providecommand \Eprint [0]{\href }%
\providecommand \doibase [0]{https://doi.org/}%
\providecommand \selectlanguage [0]{\@gobble}%
\providecommand \bibinfo  [0]{\@secondoftwo}%
\providecommand \bibfield  [0]{\@secondoftwo}%
\providecommand \translation [1]{[#1]}%
\providecommand \BibitemOpen [0]{}%
\providecommand \bibitemStop [0]{}%
\providecommand \bibitemNoStop [0]{.\EOS\space}%
\providecommand \EOS [0]{\spacefactor3000\relax}%
\providecommand \BibitemShut  [1]{\csname bibitem#1\endcsname}%
\let\auto@bib@innerbib\@empty
\bibitem [{\citenamefont {Pittman}\ \emph {et~al.}(1995)\citenamefont {Pittman}, \citenamefont {Shih}, \citenamefont {Strekalov},\ and\ \citenamefont {Sergienko}}]{PhysRevA.52.R3429}%
  \BibitemOpen
  \bibfield  {author} {\bibinfo {author} {\bibfnamefont {T.~B.}\ \bibnamefont {Pittman}}, \bibinfo {author} {\bibfnamefont {Y.~H.}\ \bibnamefont {Shih}}, \bibinfo {author} {\bibfnamefont {D.~V.}\ \bibnamefont {Strekalov}},\ and\ \bibinfo {author} {\bibfnamefont {A.~V.}\ \bibnamefont {Sergienko}},\ }\bibfield  {title} {\bibinfo {title} {Optical imaging by means of two-photon quantum entanglement},\ }\href {https://doi.org/10.1103/PhysRevA.52.R3429} {\bibfield  {journal} {\bibinfo  {journal} {Phys. Rev. A}\ }\textbf {\bibinfo {volume} {52}},\ \bibinfo {pages} {R3429} (\bibinfo {year} {1995})}\BibitemShut {NoStop}%
\bibitem [{\citenamefont {Bennink}\ \emph {et~al.}(2002)\citenamefont {Bennink}, \citenamefont {Bentley},\ and\ \citenamefont {Boyd}}]{PhysRevLett.89.113601}%
  \BibitemOpen
  \bibfield  {author} {\bibinfo {author} {\bibfnamefont {R.~S.}\ \bibnamefont {Bennink}}, \bibinfo {author} {\bibfnamefont {S.~J.}\ \bibnamefont {Bentley}},\ and\ \bibinfo {author} {\bibfnamefont {R.~W.}\ \bibnamefont {Boyd}},\ }\bibfield  {title} {\bibinfo {title} {Two-photon coincidence imaging with a classical source},\ }\href {https://doi.org/10.1103/PhysRevLett.89.113601} {\bibfield  {journal} {\bibinfo  {journal} {Phys. Rev. Lett.}\ }\textbf {\bibinfo {volume} {89}},\ \bibinfo {pages} {113601} (\bibinfo {year} {2002})}\BibitemShut {NoStop}%
\bibitem [{\citenamefont {Valencia}\ \emph {et~al.}(2005)\citenamefont {Valencia}, \citenamefont {Scarcelli}, \citenamefont {D'Angelo},\ and\ \citenamefont {Shih}}]{PhysRevLett.94.063601}%
  \BibitemOpen
  \bibfield  {author} {\bibinfo {author} {\bibfnamefont {A.}~\bibnamefont {Valencia}}, \bibinfo {author} {\bibfnamefont {G.}~\bibnamefont {Scarcelli}}, \bibinfo {author} {\bibfnamefont {M.}~\bibnamefont {D'Angelo}},\ and\ \bibinfo {author} {\bibfnamefont {Y.}~\bibnamefont {Shih}},\ }\bibfield  {title} {\bibinfo {title} {Two-photon imaging with thermal light},\ }\href {https://doi.org/10.1103/PhysRevLett.94.063601} {\bibfield  {journal} {\bibinfo  {journal} {Phys. Rev. Lett.}\ }\textbf {\bibinfo {volume} {94}},\ \bibinfo {pages} {063601} (\bibinfo {year} {2005})}\BibitemShut {NoStop}%
\bibitem [{\citenamefont {Gatti}\ \emph {et~al.}(2006)\citenamefont {Gatti}, \citenamefont {Bache}, \citenamefont {Magatti}, \citenamefont {Brambilla}, \citenamefont {Ferri},\ and\ \citenamefont {Lugiato}}]{doi:10.1080/09500340500147240}%
  \BibitemOpen
  \bibfield  {author} {\bibinfo {author} {\bibfnamefont {A.}~\bibnamefont {Gatti}}, \bibinfo {author} {\bibfnamefont {M.}~\bibnamefont {Bache}}, \bibinfo {author} {\bibfnamefont {D.}~\bibnamefont {Magatti}}, \bibinfo {author} {\bibfnamefont {E.}~\bibnamefont {Brambilla}}, \bibinfo {author} {\bibfnamefont {F.}~\bibnamefont {Ferri}},\ and\ \bibinfo {author} {\bibfnamefont {L.~A.}\ \bibnamefont {Lugiato}},\ }\bibfield  {title} {\bibinfo {title} {Coherent imaging with pseudo-thermal incoherent light},\ }\href {https://doi.org/10.1080/09500340500147240} {\bibfield  {journal} {\bibinfo  {journal} {Journal of Modern Optics}\ }\textbf {\bibinfo {volume} {53}},\ \bibinfo {pages} {739} (\bibinfo {year} {2006})}\BibitemShut {NoStop}%
\bibitem [{\citenamefont {Erkmen}\ and\ \citenamefont {Shapiro}(2010)}]{Erkmen:10}%
  \BibitemOpen
  \bibfield  {author} {\bibinfo {author} {\bibfnamefont {B.~I.}\ \bibnamefont {Erkmen}}\ and\ \bibinfo {author} {\bibfnamefont {J.~H.}\ \bibnamefont {Shapiro}},\ }\bibfield  {title} {\bibinfo {title} {Ghost imaging: from quantum to classical to computational},\ }\href {https://doi.org/10.1364/AOP.2.000405} {\bibfield  {journal} {\bibinfo  {journal} {Adv. Opt. Photon.}\ }\textbf {\bibinfo {volume} {2}},\ \bibinfo {pages} {405} (\bibinfo {year} {2010})}\BibitemShut {NoStop}%
\bibitem [{\citenamefont {Shapiro}\ and\ \citenamefont {Boyd}(2012)}]{Shapiro2012ThePO}%
  \BibitemOpen
  \bibfield  {author} {\bibinfo {author} {\bibfnamefont {J.~H.}\ \bibnamefont {Shapiro}}\ and\ \bibinfo {author} {\bibfnamefont {R.~W.}\ \bibnamefont {Boyd}},\ }\bibfield  {title} {\bibinfo {title} {The physics of ghost imaging},\ }\href {https://api.semanticscholar.org/CorpusID:254983891} {\bibfield  {journal} {\bibinfo  {journal} {Quantum Information Processing}\ }\textbf {\bibinfo {volume} {11}},\ \bibinfo {pages} {949 } (\bibinfo {year} {2012})}\BibitemShut {NoStop}%
\bibitem [{\citenamefont {Padgett}\ and\ \citenamefont {Boyd}(2017)}]{doi:10.1098/rsta.2016.0233}%
  \BibitemOpen
  \bibfield  {author} {\bibinfo {author} {\bibfnamefont {M.~J.}\ \bibnamefont {Padgett}}\ and\ \bibinfo {author} {\bibfnamefont {R.~W.}\ \bibnamefont {Boyd}},\ }\bibfield  {title} {\bibinfo {title} {An introduction to ghost imaging: quantum and classical},\ }\href {https://doi.org/10.1098/rsta.2016.0233} {\bibfield  {journal} {\bibinfo  {journal} {Philosophical Transactions of the Royal Society A: Mathematical, Physical and Engineering Sciences}\ }\textbf {\bibinfo {volume} {375}},\ \bibinfo {pages} {20160233} (\bibinfo {year} {2017})}\BibitemShut {NoStop}%
\bibitem [{\citenamefont {Simon}\ \emph {et~al.}(2017)\citenamefont {Simon}, \citenamefont {Jaeger},\ and\ \citenamefont {Sergienko}}]{Simon2017}%
  \BibitemOpen
  \bibfield  {author} {\bibinfo {author} {\bibfnamefont {D.~S.}\ \bibnamefont {Simon}}, \bibinfo {author} {\bibfnamefont {G.}~\bibnamefont {Jaeger}},\ and\ \bibinfo {author} {\bibfnamefont {A.~V.}\ \bibnamefont {Sergienko}},\ }\bibinfo {title} {Ghost imaging and related topics},\ in\ \href {https://doi.org/10.1007/978-3-319-46551-7_6} {\emph {\bibinfo {booktitle} {Quantum Metrology, Imaging, and Communication}}}\ (\bibinfo  {publisher} {Springer International Publishing},\ \bibinfo {address} {Cham},\ \bibinfo {year} {2017})\ pp.\ \bibinfo {pages} {131--158}\BibitemShut {NoStop}%
\bibitem [{\citenamefont {Gibson}\ \emph {et~al.}(2020)\citenamefont {Gibson}, \citenamefont {Johnson},\ and\ \citenamefont {Padgett}}]{Gibson:20}%
  \BibitemOpen
  \bibfield  {author} {\bibinfo {author} {\bibfnamefont {G.~M.}\ \bibnamefont {Gibson}}, \bibinfo {author} {\bibfnamefont {S.~D.}\ \bibnamefont {Johnson}},\ and\ \bibinfo {author} {\bibfnamefont {M.~J.}\ \bibnamefont {Padgett}},\ }\bibfield  {title} {\bibinfo {title} {Single-pixel imaging 12 years on: a review},\ }\href {https://doi.org/10.1364/OE.403195} {\bibfield  {journal} {\bibinfo  {journal} {Opt. Express}\ }\textbf {\bibinfo {volume} {28}},\ \bibinfo {pages} {28190} (\bibinfo {year} {2020})}\BibitemShut {NoStop}%
\bibitem [{\citenamefont {Padgett}(2023)}]{10.1117/12.2662291}%
  \BibitemOpen
  \bibfield  {author} {\bibinfo {author} {\bibfnamefont {M.}~\bibnamefont {Padgett}},\ }\bibfield  {title} {\bibinfo {title} {{Quantum imaging overview}},\ }in\ \href {https://doi.org/10.1117/12.2662291} {\emph {\bibinfo {booktitle} {Quantum Sensing, Imaging, and Precision Metrology}}},\ Vol.\ \bibinfo {volume} {12447},\ \bibinfo {editor} {edited by\ \bibinfo {editor} {\bibfnamefont {J.}~\bibnamefont {Scheuer}}\ and\ \bibinfo {editor} {\bibfnamefont {S.~M.}\ \bibnamefont {Shahriar}}},\ \bibinfo {organization} {International Society for Optics and Photonics}\ (\bibinfo  {publisher} {SPIE},\ \bibinfo {year} {2023})\ p.\ \bibinfo {pages} {1244702}\BibitemShut {NoStop}%
\bibitem [{\citenamefont {Shapiro}(2008)}]{PhysRevA.78.061802}%
  \BibitemOpen
  \bibfield  {author} {\bibinfo {author} {\bibfnamefont {J.~H.}\ \bibnamefont {Shapiro}},\ }\bibfield  {title} {\bibinfo {title} {Computational ghost imaging},\ }\href {https://doi.org/10.1103/PhysRevA.78.061802} {\bibfield  {journal} {\bibinfo  {journal} {Phys. Rev. A}\ }\textbf {\bibinfo {volume} {78}},\ \bibinfo {pages} {061802} (\bibinfo {year} {2008})}\BibitemShut {NoStop}%
\bibitem [{\citenamefont {{Edgar}}\ \emph {et~al.}(2019)\citenamefont {{Edgar}}, \citenamefont {{Gibson}},\ and\ \citenamefont {{Padgett}}}]{2019NaPho..13...13E}%
  \BibitemOpen
  \bibfield  {author} {\bibinfo {author} {\bibfnamefont {M.~P.}\ \bibnamefont {{Edgar}}}, \bibinfo {author} {\bibfnamefont {G.~M.}\ \bibnamefont {{Gibson}}},\ and\ \bibinfo {author} {\bibfnamefont {M.~J.}\ \bibnamefont {{Padgett}}},\ }\bibfield  {title} {\bibinfo {title} {{Principles and prospects for single-pixel imaging}},\ }\href {https://doi.org/10.1038/s41566-018-0300-7} {\bibfield  {journal} {\bibinfo  {journal} {Nature Photonics}\ }\textbf {\bibinfo {volume} {13}},\ \bibinfo {pages} {13} (\bibinfo {year} {2019})}\BibitemShut {NoStop}%
\bibitem [{\citenamefont {Bruschini}\ \emph {et~al.}(2019)\citenamefont {Bruschini}, \citenamefont {Homulle}, \citenamefont {Antolovic}, \citenamefont {Burri},\ and\ \citenamefont {Charbon}}]{Bruschini}%
  \BibitemOpen
  \bibfield  {author} {\bibinfo {author} {\bibfnamefont {C.}~\bibnamefont {Bruschini}}, \bibinfo {author} {\bibfnamefont {H.}~\bibnamefont {Homulle}}, \bibinfo {author} {\bibfnamefont {I.~M.}\ \bibnamefont {Antolovic}}, \bibinfo {author} {\bibfnamefont {S.}~\bibnamefont {Burri}},\ and\ \bibinfo {author} {\bibfnamefont {E.}~\bibnamefont {Charbon}},\ }\bibfield  {title} {\bibinfo {title} {Single-photon avalanche diode imagers in biophotonics: review and outlook},\ }\href {https://doi.org/10.1038/s41377-019-0191-5} {\bibfield  {journal} {\bibinfo  {journal} {Light: Science and Applications}\ }\textbf {\bibinfo {volume} {8}},\ \bibinfo {pages} {1} (\bibinfo {year} {2019})}\BibitemShut {NoStop}%
\bibitem [{\citenamefont {Zarghami}\ \emph {et~al.}(2020)\citenamefont {Zarghami}, \citenamefont {Gasparini}, \citenamefont {Parmesan}, \citenamefont {Moreno-Garcia}, \citenamefont {Stefanov}, \citenamefont {Bessire}, \citenamefont {Unternährer},\ and\ \citenamefont {Perenzoni}}]{9142240}%
  \BibitemOpen
  \bibfield  {author} {\bibinfo {author} {\bibfnamefont {M.}~\bibnamefont {Zarghami}}, \bibinfo {author} {\bibfnamefont {L.}~\bibnamefont {Gasparini}}, \bibinfo {author} {\bibfnamefont {L.}~\bibnamefont {Parmesan}}, \bibinfo {author} {\bibfnamefont {M.}~\bibnamefont {Moreno-Garcia}}, \bibinfo {author} {\bibfnamefont {A.}~\bibnamefont {Stefanov}}, \bibinfo {author} {\bibfnamefont {B.}~\bibnamefont {Bessire}}, \bibinfo {author} {\bibfnamefont {M.}~\bibnamefont {Unternährer}},\ and\ \bibinfo {author} {\bibfnamefont {M.}~\bibnamefont {Perenzoni}},\ }\bibfield  {title} {\bibinfo {title} {A 32 x 32-pixel {CMOS} imager for quantum optics with {Per-SPAD} {TDC}, 19.48$\%$ fill-factor in a 44.64-$\mu$m pitch reaching {1-MHz} observation rate},\ }\href {https://doi.org/10.1109/JSSC.2020.3005756} {\bibfield  {journal} {\bibinfo  {journal} {IEEE Journal of Solid-State Circuits}\ }\textbf {\bibinfo {volume} {55}},\ \bibinfo {pages} {2819} (\bibinfo {year} {2020})}\BibitemShut {NoStop}%
\bibitem [{\citenamefont {Morimoto}\ \emph {et~al.}(2021)\citenamefont {Morimoto}, \citenamefont {Iwata}, \citenamefont {Shinohara}, \citenamefont {Sekine}, \citenamefont {Abdelghafar}, \citenamefont {Tsuchiya}, \citenamefont {Kuroda}, \citenamefont {Tojima}, \citenamefont {Endo}, \citenamefont {Maehashi}, \citenamefont {Ota}, \citenamefont {Sasago}, \citenamefont {Maekawa}, \citenamefont {Hikosaka}, \citenamefont {Kanou}, \citenamefont {Kato}, \citenamefont {Tezuka}, \citenamefont {Yoshizaki}, \citenamefont {Ogawa}, \citenamefont {Uehira}, \citenamefont {Ehara}, \citenamefont {Inui}, \citenamefont {Matsuno}, \citenamefont {Sakurai},\ and\ \citenamefont {Ichikawa}}]{9720605}%
  \BibitemOpen
  \bibfield  {author} {\bibinfo {author} {\bibfnamefont {K.}~\bibnamefont {Morimoto}}, \bibinfo {author} {\bibfnamefont {J.}~\bibnamefont {Iwata}}, \bibinfo {author} {\bibfnamefont {M.}~\bibnamefont {Shinohara}}, \bibinfo {author} {\bibfnamefont {H.}~\bibnamefont {Sekine}}, \bibinfo {author} {\bibfnamefont {A.}~\bibnamefont {Abdelghafar}}, \bibinfo {author} {\bibfnamefont {H.}~\bibnamefont {Tsuchiya}}, \bibinfo {author} {\bibfnamefont {Y.}~\bibnamefont {Kuroda}}, \bibinfo {author} {\bibfnamefont {K.}~\bibnamefont {Tojima}}, \bibinfo {author} {\bibfnamefont {W.}~\bibnamefont {Endo}}, \bibinfo {author} {\bibfnamefont {Y.}~\bibnamefont {Maehashi}}, \bibinfo {author} {\bibfnamefont {Y.}~\bibnamefont {Ota}}, \bibinfo {author} {\bibfnamefont {T.}~\bibnamefont {Sasago}}, \bibinfo {author} {\bibfnamefont {S.}~\bibnamefont {Maekawa}}, \bibinfo {author} {\bibfnamefont {S.}~\bibnamefont {Hikosaka}}, \bibinfo {author} {\bibfnamefont {T.}~\bibnamefont {Kanou}}, \bibinfo {author} {\bibfnamefont {A.}~\bibnamefont {Kato}},
  \bibinfo {author} {\bibfnamefont {T.}~\bibnamefont {Tezuka}}, \bibinfo {author} {\bibfnamefont {S.}~\bibnamefont {Yoshizaki}}, \bibinfo {author} {\bibfnamefont {T.}~\bibnamefont {Ogawa}}, \bibinfo {author} {\bibfnamefont {K.}~\bibnamefont {Uehira}}, \bibinfo {author} {\bibfnamefont {A.}~\bibnamefont {Ehara}}, \bibinfo {author} {\bibfnamefont {F.}~\bibnamefont {Inui}}, \bibinfo {author} {\bibfnamefont {Y.}~\bibnamefont {Matsuno}}, \bibinfo {author} {\bibfnamefont {K.}~\bibnamefont {Sakurai}},\ and\ \bibinfo {author} {\bibfnamefont {T.}~\bibnamefont {Ichikawa}},\ }\bibfield  {title} {\bibinfo {title} {3.2 megapixel {3D}-stacked charge focusing {SPAD} for low-light imaging and depth sensing},\ }\href {https://doi.org/10.1109/IEDM19574.2021.9720605} {\bibfield  {journal} {\bibinfo  {journal} {2021 IEEE International Electron Devices Meeting (IEDM)}\ ,\ \bibinfo {pages} {20.2.1}} (\bibinfo {year} {2021})}\BibitemShut {NoStop}%
\bibitem [{\citenamefont {Smith}\ \emph {et~al.}(2022)\citenamefont {Smith}, \citenamefont {Rudkouskaya}, \citenamefont {Gao}, \citenamefont {Gupta}, \citenamefont {Ulku}, \citenamefont {Bruschini}, \citenamefont {Charbon}, \citenamefont {Weiss}, \citenamefont {Barroso}, \citenamefont {Intes},\ and\ \citenamefont {Michalet}}]{Smith:22}%
  \BibitemOpen
  \bibfield  {author} {\bibinfo {author} {\bibfnamefont {J.~T.}\ \bibnamefont {Smith}}, \bibinfo {author} {\bibfnamefont {A.}~\bibnamefont {Rudkouskaya}}, \bibinfo {author} {\bibfnamefont {S.}~\bibnamefont {Gao}}, \bibinfo {author} {\bibfnamefont {J.~M.}\ \bibnamefont {Gupta}}, \bibinfo {author} {\bibfnamefont {A.}~\bibnamefont {Ulku}}, \bibinfo {author} {\bibfnamefont {C.}~\bibnamefont {Bruschini}}, \bibinfo {author} {\bibfnamefont {E.}~\bibnamefont {Charbon}}, \bibinfo {author} {\bibfnamefont {S.}~\bibnamefont {Weiss}}, \bibinfo {author} {\bibfnamefont {M.}~\bibnamefont {Barroso}}, \bibinfo {author} {\bibfnamefont {X.}~\bibnamefont {Intes}},\ and\ \bibinfo {author} {\bibfnamefont {X.}~\bibnamefont {Michalet}},\ }\bibfield  {title} {\bibinfo {title} {In vitro and in vivo {NIR} fluorescence lifetime imaging with a time-gated {SPAD} camera},\ }\href {https://doi.org/10.1364/OPTICA.454790} {\bibfield  {journal} {\bibinfo  {journal} {Optica}\ }\textbf {\bibinfo {volume} {9}},\ \bibinfo {pages} {532} (\bibinfo
  {year} {2022})}\BibitemShut {NoStop}%
\bibitem [{\citenamefont {Defienne}\ \emph {et~al.}(2022)\citenamefont {Defienne}, \citenamefont {Cameron}, \citenamefont {Ndagano}, \citenamefont {Lyons}, \citenamefont {Reichert}, \citenamefont {Zhao}, \citenamefont {Harvey}, \citenamefont {Charbon}, \citenamefont {Fleischer},\ and\ \citenamefont {Faccio}}]{defienne}%
  \BibitemOpen
  \bibfield  {author} {\bibinfo {author} {\bibfnamefont {H.}~\bibnamefont {Defienne}}, \bibinfo {author} {\bibfnamefont {P.}~\bibnamefont {Cameron}}, \bibinfo {author} {\bibfnamefont {B.}~\bibnamefont {Ndagano}}, \bibinfo {author} {\bibfnamefont {A.}~\bibnamefont {Lyons}}, \bibinfo {author} {\bibfnamefont {M.}~\bibnamefont {Reichert}}, \bibinfo {author} {\bibfnamefont {J.}~\bibnamefont {Zhao}}, \bibinfo {author} {\bibfnamefont {A.}~\bibnamefont {Harvey}}, \bibinfo {author} {\bibfnamefont {E.}~\bibnamefont {Charbon}}, \bibinfo {author} {\bibfnamefont {J.}~\bibnamefont {Fleischer}},\ and\ \bibinfo {author} {\bibfnamefont {D.}~\bibnamefont {Faccio}},\ }\bibfield  {title} {\bibinfo {title} {Pixel super-resolution with spatially entangled photons},\ }\href {https://doi.org/10.1038/s41467-022-31052-6} {\bibfield  {journal} {\bibinfo  {journal} {Nature Communications}\ }\textbf {\bibinfo {volume} {13}} (\bibinfo {year} {2022})}\BibitemShut {NoStop}%
\bibitem [{\citenamefont {Untern\"{a}hrer}\ \emph {et~al.}(2018)\citenamefont {Untern\"{a}hrer}, \citenamefont {Bessire}, \citenamefont {Gasparini}, \citenamefont {Perenzoni},\ and\ \citenamefont {Stefanov}}]{Unternahrer:18}%
  \BibitemOpen
  \bibfield  {author} {\bibinfo {author} {\bibfnamefont {M.}~\bibnamefont {Untern\"{a}hrer}}, \bibinfo {author} {\bibfnamefont {B.}~\bibnamefont {Bessire}}, \bibinfo {author} {\bibfnamefont {L.}~\bibnamefont {Gasparini}}, \bibinfo {author} {\bibfnamefont {M.}~\bibnamefont {Perenzoni}},\ and\ \bibinfo {author} {\bibfnamefont {A.}~\bibnamefont {Stefanov}},\ }\bibfield  {title} {\bibinfo {title} {Super-resolution quantum imaging at the heisenberg limit},\ }\href {https://doi.org/10.1364/OPTICA.5.001150} {\bibfield  {journal} {\bibinfo  {journal} {Optica}\ }\textbf {\bibinfo {volume} {5}},\ \bibinfo {pages} {1150} (\bibinfo {year} {2018})}\BibitemShut {NoStop}%
\bibitem [{\citenamefont {Eckmann}\ \emph {et~al.}(2020)\citenamefont {Eckmann}, \citenamefont {Bessire}, \citenamefont {Untern\"{a}hrer}, \citenamefont {Gasparini}, \citenamefont {Perenzoni},\ and\ \citenamefont {Stefanov}}]{Eckmann:20}%
  \BibitemOpen
  \bibfield  {author} {\bibinfo {author} {\bibfnamefont {B.}~\bibnamefont {Eckmann}}, \bibinfo {author} {\bibfnamefont {B.}~\bibnamefont {Bessire}}, \bibinfo {author} {\bibfnamefont {M.}~\bibnamefont {Untern\"{a}hrer}}, \bibinfo {author} {\bibfnamefont {L.}~\bibnamefont {Gasparini}}, \bibinfo {author} {\bibfnamefont {M.}~\bibnamefont {Perenzoni}},\ and\ \bibinfo {author} {\bibfnamefont {A.}~\bibnamefont {Stefanov}},\ }\bibfield  {title} {\bibinfo {title} {Characterization of space-momentum entangled photons with a time resolving {CMOS} {SPAD} array},\ }\href {https://doi.org/10.1364/OE.401260} {\bibfield  {journal} {\bibinfo  {journal} {Opt. Express}\ }\textbf {\bibinfo {volume} {28}},\ \bibinfo {pages} {31553} (\bibinfo {year} {2020})}\BibitemShut {NoStop}%
\bibitem [{\citenamefont {Abbattista}\ \emph {et~al.}(2021)\citenamefont {Abbattista}, \citenamefont {Amoruso}, \citenamefont {Burri}, \citenamefont {Charbon}, \citenamefont {Di~Lena}, \citenamefont {Garuccio}, \citenamefont {Giannella}, \citenamefont {Hradil}, \citenamefont {Iacobellis}, \citenamefont {Massaro}, \citenamefont {Mos}, \citenamefont {Motka}, \citenamefont {Paúr}, \citenamefont {Pepe}, \citenamefont {Peterek}, \citenamefont {Petrelli}, \citenamefont {Řeháček}, \citenamefont {Santoro}, \citenamefont {Scattarella}, \citenamefont {Ulku}, \citenamefont {Vasiukov}, \citenamefont {Wayne}, \citenamefont {Bruschini}, \citenamefont {D’Angelo}, \citenamefont {Ieronymaki},\ and\ \citenamefont {Stoklasa}}]{app11146414}%
  \BibitemOpen
  \bibfield  {author} {\bibinfo {author} {\bibfnamefont {C.}~\bibnamefont {Abbattista}}, \bibinfo {author} {\bibfnamefont {L.}~\bibnamefont {Amoruso}}, \bibinfo {author} {\bibfnamefont {S.}~\bibnamefont {Burri}}, \bibinfo {author} {\bibfnamefont {E.}~\bibnamefont {Charbon}}, \bibinfo {author} {\bibfnamefont {F.}~\bibnamefont {Di~Lena}}, \bibinfo {author} {\bibfnamefont {A.}~\bibnamefont {Garuccio}}, \bibinfo {author} {\bibfnamefont {D.}~\bibnamefont {Giannella}}, \bibinfo {author} {\bibfnamefont {Z.}~\bibnamefont {Hradil}}, \bibinfo {author} {\bibfnamefont {M.}~\bibnamefont {Iacobellis}}, \bibinfo {author} {\bibfnamefont {G.}~\bibnamefont {Massaro}}, \bibinfo {author} {\bibfnamefont {P.}~\bibnamefont {Mos}}, \bibinfo {author} {\bibfnamefont {L.}~\bibnamefont {Motka}}, \bibinfo {author} {\bibfnamefont {M.}~\bibnamefont {Paúr}}, \bibinfo {author} {\bibfnamefont {F.~V.}\ \bibnamefont {Pepe}}, \bibinfo {author} {\bibfnamefont {M.}~\bibnamefont {Peterek}}, \bibinfo {author} {\bibfnamefont {I.}~\bibnamefont
  {Petrelli}}, \bibinfo {author} {\bibfnamefont {J.}~\bibnamefont {Řeháček}}, \bibinfo {author} {\bibfnamefont {F.}~\bibnamefont {Santoro}}, \bibinfo {author} {\bibfnamefont {F.}~\bibnamefont {Scattarella}}, \bibinfo {author} {\bibfnamefont {A.}~\bibnamefont {Ulku}}, \bibinfo {author} {\bibfnamefont {S.}~\bibnamefont {Vasiukov}}, \bibinfo {author} {\bibfnamefont {M.}~\bibnamefont {Wayne}}, \bibinfo {author} {\bibfnamefont {C.}~\bibnamefont {Bruschini}}, \bibinfo {author} {\bibfnamefont {M.}~\bibnamefont {D’Angelo}}, \bibinfo {author} {\bibfnamefont {M.}~\bibnamefont {Ieronymaki}},\ and\ \bibinfo {author} {\bibfnamefont {B.}~\bibnamefont {Stoklasa}},\ }\bibfield  {title} {\bibinfo {title} {Towards quantum {3D} imaging devices},\ }\bibfield  {journal} {\bibinfo  {journal} {Applied Sciences}\ }\textbf {\bibinfo {volume} {11}},\ \href {https://doi.org/10.3390/app11146414} {10.3390/app11146414} (\bibinfo {year} {2021})\BibitemShut {NoStop}%
\bibitem [{\citenamefont {Ndagano}\ \emph {et~al.}(2022)\citenamefont {Ndagano}, \citenamefont {Defienne}, \citenamefont {Branford}, \citenamefont {Shah}, \citenamefont {Lyons}, \citenamefont {Westerberg}, \citenamefont {Gauger},\ and\ \citenamefont {Faccio}}]{ndagano}%
  \BibitemOpen
  \bibfield  {author} {\bibinfo {author} {\bibfnamefont {B.}~\bibnamefont {Ndagano}}, \bibinfo {author} {\bibfnamefont {H.}~\bibnamefont {Defienne}}, \bibinfo {author} {\bibfnamefont {D.}~\bibnamefont {Branford}}, \bibinfo {author} {\bibfnamefont {Y.}~\bibnamefont {Shah}}, \bibinfo {author} {\bibfnamefont {A.}~\bibnamefont {Lyons}}, \bibinfo {author} {\bibfnamefont {N.}~\bibnamefont {Westerberg}}, \bibinfo {author} {\bibfnamefont {E.}~\bibnamefont {Gauger}},\ and\ \bibinfo {author} {\bibfnamefont {D.}~\bibnamefont {Faccio}},\ }\bibfield  {title} {\bibinfo {title} {Quantum microscopy based on {Hong-Ou-Mandel} interference},\ }\href {https://doi.org/10.1038/s41566-022-00980-6} {\bibfield  {journal} {\bibinfo  {journal} {Nature Photonics}\ }\textbf {\bibinfo {volume} {16}},\ \bibinfo {pages} {384} (\bibinfo {year} {2022})}\BibitemShut {NoStop}%
\bibitem [{\citenamefont {Zhao}\ \emph {et~al.}(2022)\citenamefont {Zhao}, \citenamefont {Lyons}, \citenamefont {Ulku}, \citenamefont {Defienne}, \citenamefont {Faccio},\ and\ \citenamefont {Charbon}}]{Zhao:22}%
  \BibitemOpen
  \bibfield  {author} {\bibinfo {author} {\bibfnamefont {J.}~\bibnamefont {Zhao}}, \bibinfo {author} {\bibfnamefont {A.}~\bibnamefont {Lyons}}, \bibinfo {author} {\bibfnamefont {A.~C.}\ \bibnamefont {Ulku}}, \bibinfo {author} {\bibfnamefont {H.}~\bibnamefont {Defienne}}, \bibinfo {author} {\bibfnamefont {D.}~\bibnamefont {Faccio}},\ and\ \bibinfo {author} {\bibfnamefont {E.}~\bibnamefont {Charbon}},\ }\bibfield  {title} {\bibinfo {title} {Light detection and ranging with entangled photons},\ }\href {https://doi.org/10.1364/OE.435898} {\bibfield  {journal} {\bibinfo  {journal} {Opt. Express}\ }\textbf {\bibinfo {volume} {30}},\ \bibinfo {pages} {3675} (\bibinfo {year} {2022})}\BibitemShut {NoStop}%
\bibitem [{\citenamefont {Braga}\ \emph {et~al.}(2014)\citenamefont {Braga}, \citenamefont {Gasparini}, \citenamefont {Grant}, \citenamefont {Henderson}, \citenamefont {Massari}, \citenamefont {Perenzoni}, \citenamefont {Stoppa},\ and\ \citenamefont {Walker}}]{6642135}%
  \BibitemOpen
  \bibfield  {author} {\bibinfo {author} {\bibfnamefont {L.~H.~C.}\ \bibnamefont {Braga}}, \bibinfo {author} {\bibfnamefont {L.}~\bibnamefont {Gasparini}}, \bibinfo {author} {\bibfnamefont {L.}~\bibnamefont {Grant}}, \bibinfo {author} {\bibfnamefont {R.~K.}\ \bibnamefont {Henderson}}, \bibinfo {author} {\bibfnamefont {N.}~\bibnamefont {Massari}}, \bibinfo {author} {\bibfnamefont {M.}~\bibnamefont {Perenzoni}}, \bibinfo {author} {\bibfnamefont {D.}~\bibnamefont {Stoppa}},\ and\ \bibinfo {author} {\bibfnamefont {R.}~\bibnamefont {Walker}},\ }\bibfield  {title} {\bibinfo {title} {A fully digital 8$\,\times\,$16 {SiPM} array for {PET} applications with per-pixel {TDCs} and real-time energy output},\ }\href {https://doi.org/10.1109/JSSC.2013.2284351} {\bibfield  {journal} {\bibinfo  {journal} {IEEE Journal of Solid-State Circuits}\ }\textbf {\bibinfo {volume} {49}},\ \bibinfo {pages} {301} (\bibinfo {year} {2014})}\BibitemShut {NoStop}%
\bibitem [{\citenamefont {Carimatto}\ \emph {et~al.}(2018)\citenamefont {Carimatto}, \citenamefont {Ulku}, \citenamefont {Lindner}, \citenamefont {Gros-Daillon}, \citenamefont {Rae}, \citenamefont {Pellegrini},\ and\ \citenamefont {Charbon}}]{8700491}%
  \BibitemOpen
  \bibfield  {author} {\bibinfo {author} {\bibfnamefont {A.}~\bibnamefont {Carimatto}}, \bibinfo {author} {\bibfnamefont {A.}~\bibnamefont {Ulku}}, \bibinfo {author} {\bibfnamefont {S.}~\bibnamefont {Lindner}}, \bibinfo {author} {\bibfnamefont {E.}~\bibnamefont {Gros-Daillon}}, \bibinfo {author} {\bibfnamefont {B.}~\bibnamefont {Rae}}, \bibinfo {author} {\bibfnamefont {S.}~\bibnamefont {Pellegrini}},\ and\ \bibinfo {author} {\bibfnamefont {E.}~\bibnamefont {Charbon}},\ }\bibfield  {title} {\bibinfo {title} {Multipurpose, fully integrated 128 $\times$ 128 event-driven {MD-SiPM} with 512 16-bit {TDCs} with 45-ps {LSB} and 20-ns gating in 40-nm {CMOS} technology},\ }\href {https://doi.org/10.1109/LSSC.2019.2911043} {\bibfield  {journal} {\bibinfo  {journal} {IEEE Solid-State Circuits Letters}\ }\textbf {\bibinfo {volume} {1}},\ \bibinfo {pages} {241} (\bibinfo {year} {2018})}\BibitemShut {NoStop}%
\bibitem [{\citenamefont {Xu}\ \emph {et~al.}(2017)\citenamefont {Xu}, \citenamefont {Pancheri}, \citenamefont {Betta},\ and\ \citenamefont {Stoppa}}]{Xu:17}%
  \BibitemOpen
  \bibfield  {author} {\bibinfo {author} {\bibfnamefont {H.}~\bibnamefont {Xu}}, \bibinfo {author} {\bibfnamefont {L.}~\bibnamefont {Pancheri}}, \bibinfo {author} {\bibfnamefont {G.-F.~D.}\ \bibnamefont {Betta}},\ and\ \bibinfo {author} {\bibfnamefont {D.}~\bibnamefont {Stoppa}},\ }\bibfield  {title} {\bibinfo {title} {Design and characterization of a p$+$/n-well {SPAD} array in 150nm {CMOS} process},\ }\href {https://doi.org/10.1364/OE.25.012765} {\bibfield  {journal} {\bibinfo  {journal} {Opt. Express}\ }\textbf {\bibinfo {volume} {25}},\ \bibinfo {pages} {12765} (\bibinfo {year} {2017})}\BibitemShut {NoStop}%
\bibitem [{\citenamefont {Magnitskiy}\ \emph {et~al.}(2022)\citenamefont {Magnitskiy}, \citenamefont {Agapov},\ and\ \citenamefont {Chirkin}}]{Magnitskiy:22}%
  \BibitemOpen
  \bibfield  {author} {\bibinfo {author} {\bibfnamefont {S.}~\bibnamefont {Magnitskiy}}, \bibinfo {author} {\bibfnamefont {D.}~\bibnamefont {Agapov}},\ and\ \bibinfo {author} {\bibfnamefont {A.}~\bibnamefont {Chirkin}},\ }\bibfield  {title} {\bibinfo {title} {Quantum ghost polarimetry with entangled photons},\ }\href {https://doi.org/10.1364/OL.450206} {\bibfield  {journal} {\bibinfo  {journal} {Opt. Lett.}\ }\textbf {\bibinfo {volume} {47}},\ \bibinfo {pages} {754} (\bibinfo {year} {2022})}\BibitemShut {NoStop}%
\bibitem [{\citenamefont {Magnitskiy}\ \emph {et~al.}(2020)\citenamefont {Magnitskiy}, \citenamefont {Agapov},\ and\ \citenamefont {Chirkin}}]{Magnitskiy:20}%
  \BibitemOpen
  \bibfield  {author} {\bibinfo {author} {\bibfnamefont {S.}~\bibnamefont {Magnitskiy}}, \bibinfo {author} {\bibfnamefont {D.}~\bibnamefont {Agapov}},\ and\ \bibinfo {author} {\bibfnamefont {A.}~\bibnamefont {Chirkin}},\ }\bibfield  {title} {\bibinfo {title} {Ghost polarimetry with unpolarized pseudo-thermal light},\ }\href {https://doi.org/10.1364/OL.387234} {\bibfield  {journal} {\bibinfo  {journal} {Opt. Lett.}\ }\textbf {\bibinfo {volume} {45}},\ \bibinfo {pages} {3641} (\bibinfo {year} {2020})}\BibitemShut {NoStop}%
\bibitem [{\citenamefont {{Magnitskiy}}\ \emph {et~al.}(2021)\citenamefont {{Magnitskiy}}, \citenamefont {{Agapov}}, \citenamefont {{Belovolov}}, \citenamefont {{Gostev}}, \citenamefont {{Frolovtsev}},\ and\ \citenamefont {{Chirkin}}}]{Magnitskiy:21}%
  \BibitemOpen
  \bibfield  {author} {\bibinfo {author} {\bibfnamefont {S.~A.}\ \bibnamefont {{Magnitskiy}}}, \bibinfo {author} {\bibfnamefont {D.~P.}\ \bibnamefont {{Agapov}}}, \bibinfo {author} {\bibfnamefont {I.~A.}\ \bibnamefont {{Belovolov}}}, \bibinfo {author} {\bibfnamefont {P.~P.}\ \bibnamefont {{Gostev}}}, \bibinfo {author} {\bibfnamefont {D.~N.}\ \bibnamefont {{Frolovtsev}}},\ and\ \bibinfo {author} {\bibfnamefont {A.~S.}\ \bibnamefont {{Chirkin}}},\ }\bibfield  {title} {\bibinfo {title} {{Ghost Polarimetry in Classical and Quantum Light}},\ }\href {https://doi.org/10.3103/S0027134921060060} {\bibfield  {journal} {\bibinfo  {journal} {Moscow University Physics Bulletin}\ }\textbf {\bibinfo {volume} {76}},\ \bibinfo {pages} {424} (\bibinfo {year} {2021})}\BibitemShut {NoStop}%
\bibitem [{Tar()}]{Target}%
  \BibitemOpen
  \href@noop {} {}\bibinfo {howpublished} {\url{https://www.savazzi.net/photography/1951usaf.html}}\BibitemShut {NoStop}%
\bibitem [{\citenamefont {Pitsch}\ \emph {et~al.}(2023)\citenamefont {Pitsch}, \citenamefont {Walter}, \citenamefont {Gasparini}, \citenamefont {B\"{u}rsing},\ and\ \citenamefont {Eichhorn}}]{Pitsch:23}%
  \BibitemOpen
  \bibfield  {author} {\bibinfo {author} {\bibfnamefont {C.}~\bibnamefont {Pitsch}}, \bibinfo {author} {\bibfnamefont {D.}~\bibnamefont {Walter}}, \bibinfo {author} {\bibfnamefont {L.}~\bibnamefont {Gasparini}}, \bibinfo {author} {\bibfnamefont {H.}~\bibnamefont {B\"{u}rsing}},\ and\ \bibinfo {author} {\bibfnamefont {M.}~\bibnamefont {Eichhorn}},\ }\bibfield  {title} {\bibinfo {title} {{3D} quantum ghost imaging},\ }\href {https://doi.org/10.1364/AO.492208} {\bibfield  {journal} {\bibinfo  {journal} {Appl. Opt.}\ }\textbf {\bibinfo {volume} {62}},\ \bibinfo {pages} {6275} (\bibinfo {year} {2023})}\BibitemShut {NoStop}%
\bibitem [{\citenamefont {Gili}\ \emph {et~al.}(2023)\citenamefont {Gili}, \citenamefont {Dupish}, \citenamefont {Vega}, \citenamefont {Gandola}, \citenamefont {Manuzzato}, \citenamefont {Perenzoni}, \citenamefont {Gasparini}, \citenamefont {Pertsch},\ and\ \citenamefont {Setzpfandt}}]{Gili:23}%
  \BibitemOpen
  \bibfield  {author} {\bibinfo {author} {\bibfnamefont {V.~F.}\ \bibnamefont {Gili}}, \bibinfo {author} {\bibfnamefont {D.}~\bibnamefont {Dupish}}, \bibinfo {author} {\bibfnamefont {A.}~\bibnamefont {Vega}}, \bibinfo {author} {\bibfnamefont {M.}~\bibnamefont {Gandola}}, \bibinfo {author} {\bibfnamefont {E.}~\bibnamefont {Manuzzato}}, \bibinfo {author} {\bibfnamefont {M.}~\bibnamefont {Perenzoni}}, \bibinfo {author} {\bibfnamefont {L.}~\bibnamefont {Gasparini}}, \bibinfo {author} {\bibfnamefont {T.}~\bibnamefont {Pertsch}},\ and\ \bibinfo {author} {\bibfnamefont {F.}~\bibnamefont {Setzpfandt}},\ }\bibfield  {title} {\bibinfo {title} {Quantum ghost imaging based on a ``looking back'' {2D} {SPAD} array},\ }\href {https://doi.org/10.1364/AO.487084} {\bibfield  {journal} {\bibinfo  {journal} {Appl. Opt.}\ }\textbf {\bibinfo {volume} {62}},\ \bibinfo {pages} {3093} (\bibinfo {year} {2023})}\BibitemShut {NoStop}%
\bibitem [{\citenamefont {{Starovoitov}}\ \emph {et~al.}(2023)\citenamefont {{Starovoitov}}, \citenamefont {{Chizhevsky}}, \citenamefont {{Horoshko}},\ and\ \citenamefont {{Kilin}}}]{2023JApSp..90..377S}%
  \BibitemOpen
  \bibfield  {author} {\bibinfo {author} {\bibfnamefont {V.~S.}\ \bibnamefont {{Starovoitov}}}, \bibinfo {author} {\bibfnamefont {V.~N.}\ \bibnamefont {{Chizhevsky}}}, \bibinfo {author} {\bibfnamefont {D.~B.}\ \bibnamefont {{Horoshko}}},\ and\ \bibinfo {author} {\bibfnamefont {S.~Y.}\ \bibnamefont {{Kilin}}},\ }\bibfield  {title} {\bibinfo {title} {{Spatio-Temporal Correlations of Photons from a Pseudo-Thermal Source}},\ }\href {https://doi.org/10.1007/s10812-023-01544-4} {\bibfield  {journal} {\bibinfo  {journal} {Journal of Applied Spectroscopy}\ }\textbf {\bibinfo {volume} {90}},\ \bibinfo {pages} {377} (\bibinfo {year} {2023})}\BibitemShut {NoStop}%
\bibitem [{\citenamefont {Wang}\ \emph {et~al.}(2004)\citenamefont {Wang}, \citenamefont {Bovik}, \citenamefont {Sheikh},\ and\ \citenamefont {Simoncelli}}]{wang2004image}%
  \BibitemOpen
  \bibfield  {author} {\bibinfo {author} {\bibfnamefont {Z.}~\bibnamefont {Wang}}, \bibinfo {author} {\bibfnamefont {A.~C.}\ \bibnamefont {Bovik}}, \bibinfo {author} {\bibfnamefont {H.~R.}\ \bibnamefont {Sheikh}},\ and\ \bibinfo {author} {\bibfnamefont {E.~P.}\ \bibnamefont {Simoncelli}},\ }\bibfield  {title} {\bibinfo {title} {Image quality assessment: from error visibility to structural similarity},\ }\href@noop {} {\bibfield  {journal} {\bibinfo  {journal} {IEEE transactions on image processing}\ }\textbf {\bibinfo {volume} {13}},\ \bibinfo {pages} {600} (\bibinfo {year} {2004})}\BibitemShut {NoStop}%
\bibitem [{\citenamefont {{Ryczkowski}}\ \emph {et~al.}(2016)\citenamefont {{Ryczkowski}}, \citenamefont {{Barbier}}, \citenamefont {{Friberg}}, \citenamefont {{Dudley}},\ and\ \citenamefont {{Genty}}}]{2016NaPho..10..167R}%
  \BibitemOpen
  \bibfield  {author} {\bibinfo {author} {\bibfnamefont {P.}~\bibnamefont {{Ryczkowski}}}, \bibinfo {author} {\bibfnamefont {M.}~\bibnamefont {{Barbier}}}, \bibinfo {author} {\bibfnamefont {A.~T.}\ \bibnamefont {{Friberg}}}, \bibinfo {author} {\bibfnamefont {J.~M.}\ \bibnamefont {{Dudley}}},\ and\ \bibinfo {author} {\bibfnamefont {G.}~\bibnamefont {{Genty}}},\ }\bibfield  {title} {\bibinfo {title} {{Ghost imaging in the time domain}},\ }\href {https://doi.org/10.1038/nphoton.2015.274} {\bibfield  {journal} {\bibinfo  {journal} {Nature Photonics}\ }\textbf {\bibinfo {volume} {10}},\ \bibinfo {pages} {167} (\bibinfo {year} {2016})}\BibitemShut {NoStop}%
\bibitem [{\citenamefont {Devaux}\ \emph {et~al.}(2016)\citenamefont {Devaux}, \citenamefont {Moreau}, \citenamefont {Denis},\ and\ \citenamefont {Lantz}}]{Devaux:16}%
  \BibitemOpen
  \bibfield  {author} {\bibinfo {author} {\bibfnamefont {F.}~\bibnamefont {Devaux}}, \bibinfo {author} {\bibfnamefont {P.-A.}\ \bibnamefont {Moreau}}, \bibinfo {author} {\bibfnamefont {S.}~\bibnamefont {Denis}},\ and\ \bibinfo {author} {\bibfnamefont {E.}~\bibnamefont {Lantz}},\ }\bibfield  {title} {\bibinfo {title} {Computational temporal ghost imaging},\ }\href {https://doi.org/10.1364/OPTICA.3.000698} {\bibfield  {journal} {\bibinfo  {journal} {Optica}\ }\textbf {\bibinfo {volume} {3}},\ \bibinfo {pages} {698} (\bibinfo {year} {2016})}\BibitemShut {NoStop}%
\bibitem [{\citenamefont {Horoshko}(2023)}]{Horoshko:23}%
  \BibitemOpen
  \bibfield  {author} {\bibinfo {author} {\bibfnamefont {D.~B.}\ \bibnamefont {Horoshko}},\ }\bibfield  {title} {\bibinfo {title} {Time-to-space ghost imaging},\ }\href {https://doi.org/10.1364/OL.487394} {\bibfield  {journal} {\bibinfo  {journal} {Opt. Lett.}\ }\textbf {\bibinfo {volume} {48}},\ \bibinfo {pages} {3247} (\bibinfo {year} {2023})}\BibitemShut {NoStop}%
\bibitem [{\citenamefont {Moreau}\ \emph {et~al.}(2018)\citenamefont {Moreau}, \citenamefont {Toninelli}, \citenamefont {Morris}, \citenamefont {Aspden}, \citenamefont {Gregory}, \citenamefont {Spalding}, \citenamefont {Boyd},\ and\ \citenamefont {Padgett}}]{Moreau:18}%
  \BibitemOpen
  \bibfield  {author} {\bibinfo {author} {\bibfnamefont {P.-A.}\ \bibnamefont {Moreau}}, \bibinfo {author} {\bibfnamefont {E.}~\bibnamefont {Toninelli}}, \bibinfo {author} {\bibfnamefont {P.~A.}\ \bibnamefont {Morris}}, \bibinfo {author} {\bibfnamefont {R.~S.}\ \bibnamefont {Aspden}}, \bibinfo {author} {\bibfnamefont {T.}~\bibnamefont {Gregory}}, \bibinfo {author} {\bibfnamefont {G.}~\bibnamefont {Spalding}}, \bibinfo {author} {\bibfnamefont {R.~W.}\ \bibnamefont {Boyd}},\ and\ \bibinfo {author} {\bibfnamefont {M.~J.}\ \bibnamefont {Padgett}},\ }\bibfield  {title} {\bibinfo {title} {Resolution limits of quantum ghost imaging},\ }\href {https://doi.org/10.1364/OE.26.007528} {\bibfield  {journal} {\bibinfo  {journal} {Opt. Express}\ }\textbf {\bibinfo {volume} {26}},\ \bibinfo {pages} {7528} (\bibinfo {year} {2018})}\BibitemShut {NoStop}%
\end{thebibliography}%

\end{document}